\documentclass[11pt]{amsart}
\usepackage{amsbsy,amssymb,amscd,amsfonts,latexsym,amstext,delarray,
amsmath,graphicx} 

\newtheorem{thm}{Theorem}[section]
\newtheorem{prop}[thm]{Proposition}
\newtheorem{cor}[thm]{Corollary}
\newtheorem{lem}[thm]{Lemma}

\newtheorem{rem}[thm]{Remark}

\numberwithin{equation}{section}

\def\bC{{\mathbb C}}

\def\bN{{\mathbb N}}

\def\bR{{\mathbb R}}

\def\C{{\mathbb C}}

\def\N{{\mathbb N}}

\def\Z{{\mathbb Z}}
\def\R{{\mathbb R}}

\def\cA{{\mathcal A}}

\def\cD{{\mathcal D}}
\def\cE{{\mathcal E}}
\def\cF{{\mathcal F}}

\def\cH{{\mathcal H}}

\def\cL{{\mathcal L}}

\def\cP{{\mathcal P}}

\def\cS{{\mathcal S}}

\def\cV{{\mathcal V}}
\def\cW{{\mathcal W}}

\def\Cl{{\rm Cl}}

\def\End{{\rm End}}

\def\GL{{\rm GL}}

\def\Hom{{\rm Hom}}
\def\id{{\rm id}}

\def\Tr{{\rm Tr}}
\def\tr{{\rm tr}}

\def\Vol{{\rm Vol}}

\def\cancel#1#2{\ooalign{$\hfil#1\mkern1mu/\hfil$\crcr$#1#2$}}
\def\Dirac{\mathpalette\cancel D}

\def\cutint{{\int \!\!\!\!\!\! -}}

\title[Coupling to matter, spectral action, cosmic topology]
{Coupling of gravity to matter, 
spectral action and cosmic topology}
\author{Branimir \'Ca\'ci\'c, Matilde Marcolli, Kevin Teh}
\address{Department of Mathematics  \\
California Institute of Technology \\ 
Pasadena, CA 91125, USA}
\email{branimir\@@caltech.edu}
\email{matilde\@@caltech.edu}
\email{teh\@@caltech.edu}

\begin{document}

\begin{abstract}
We consider a model of modified gravity based on the spectral action
functional, for a cosmic topology given by a spherical space form,
and the associated slow-roll inflation scenario. We
consider then the coupling of gravity to matter determined by an 
almost-commutative geometry over the spherical space form.  
We show that this produces a multiplicative shift of the 
amplitude of the power spectra for the density fluctuations and the gravitational
waves, by a multiplicative factor equal to the total number of fermions in 
the matter sector of the model.  We obtain the result by an explicit
nonperturbative computation, based on the Poisson summation formula
and the spectra of twisted Dirac operators on spherical space forms,
as well as, for more general spacetime manifolds, 
using a heat-kernel computation.
\end{abstract}

\maketitle

\tableofcontents

\section{Introduction}

Models of gravity coupled to matter based on Noncommutative Geometry
are usually obtained (see \cite{BroSuij}, \cite{ChCo}, \cite{ChCoMa}, \cite{CoSM})
by considering an underlying geometry given by a product $X\times F$
of an ordinary 4-dimensional (Riemannian compact) spacetime manifold and a finite 
noncommutative space $F$. 

The main purpose of the paper is to show how the slow-roll
inflation potential derived in \cite{MaPieTeh}, \cite{MaPieTeh2} is affected
by the presence of the matter content and the almost-commutative geometry.
We first consider the case of spherical space forms using the Poisson summation formula technique and the
nonperturbative calculation of the spectral action, and then the case of more general spacetime manifolds using a nonperturbative heat kernel
argument, to show that the amplitude of the slow-roll potential is affected by
a multiplicative factor $N$ equal to the dimension of the representation,
that is, to the total number of fermions in the theory. 

In terms of identifying specific properties of this modified gravity model
based on the spectral action, which set it apart from other models, the
dependence of the amplitude of the inflation potential on the number of
fermions in the particle physics sector of the model is so far the most
striking feature that distinguishes it from other known slow-roll inflation models.

\subsection{Spectral triples}

Noncommutative spaces are described, in this
context, as a generalization of Riemannian manifolds, via the formalism of
{\em spectral triples} introduced in \cite{CoS3}. An ordinary Riemannian
spin manifold $X$ is identified with the spectral triple $(C^\infty(X), L^2(X,S),\Dirac)$,
with the algebra of smooth functions acting as multiplication operators on 
the Hilbert space of square integrable spinors, and the Riemannian metric reconstructed
from the Dirac operator $\Dirac$. 

More generally, for a noncommutative space, a spectral triple is a similar set
$(\cA,\cH,D)$ consisting of a $*$-algebra represented by bounded operators
on a Hilbert space $\cH$, and a self-adjoint operator $D$ with compact resolvent
acting on $\cH$ with a dense domain and such that the commutators $[D,a]$
extend to bounded operators on all of $\cH$. A {\em finite} noncommutative
space is one for which the algebra $\cA$ and Hilbert space $\cH$ are finite dimensional.

A recent powerful reconstruction theorem (\cite{Co-reco}, see also \cite{ReVa})
shows that commutative spectral triples that satisfy certain natural axioms,
related to properties such as orientability, are spectral
triples of smooth Riemannian manifolds in the sense mentioned above.

In the models of gravity coupled to matter, the choice of the finite geometry 
$F=(\cA_F,\cH_F,D_F)$ determines the field content of the particle
physics model. 
As shown in \cite{ChCoMa}, the coordinates on the moduli 
space of possible Dirac operators $D_F$ on the finite geometry $(\cA_F,\cH_F)$ 
specify the Yukawa parameters (Dirac and Majorana masses and mixing angles)
for the particles. A classification of the moduli spaces of Dirac operators on
general finite geometries was given in \cite{Cacic1}. 

\subsection{The spectral action}

One obtains then a  theory of (modified) gravity coupled to matter by taking as an action
functional the spectral action on the noncommutative space $X\times F$, considered
as the product of the spectral triples $(C^\infty(X), L^2(X,S),\Dirac)$ and 
$(\cA_F,\cH_F,D_F)$. 

The {\em spectral action functional} introduced in \cite{ChCo} 
is a function of the spectrum of the
Dirac operator on a spectral triple, given by summing over the spectrum 
with a cutoff function. Namely, the spectral action functional is defined as
$\Tr(f(D/\Lambda))$, where $\Lambda$ is an energy scale, $D$ is the Dirac
operator of the spectral triple, and $f$ is a smooth approximation to a 
cutoff function.  As shown in \cite{ChCo} this action functional has an
asymptotic expansion at high energies  $\Lambda$ of the form
\begin{equation}\label{spactexpand}
\Tr(f(D/\Lambda)) \sim \sum_{k \in {\rm DimSp}} f_k \Lambda^k \cutint | D |^{-k}
+ f(0) \zeta_D(0) + o(1), 
\end{equation}
where the $f_k$ are the momenta $f_k=\int_0^\infty f(v) v^{k-1} dv$
of the test function $f$, for $k$ a non-negative integer in the dimension
spectrum of $D$ (the set of poles of the zeta functions 
$\zeta_{a,D}(s)=\Tr(a |D|^{-s})$) and the term $\cutint |D|^{-k}$ given by the
residue at $k$ of the zeta function $\zeta_D (s)$.  

These terms in the asymptotic expansion of the spectral action can be
computed explicitly: for a suitable choice of the finite geometry spectral triple
$(\cA_F,\cH_F,D_F)$ as in \cite{ChCoMa}, they recover all the bosonic 
terms in the Lagrangian of the Standard Model (with additional right-handed
neutrinos with Majorana mass terms) and gravitational terms including
the Einstein--Hilbert action, a cosmological term, and conformal gravity terms,
see also Chapter 1 of \cite{CoMa}. For a different choice of the finite geometry,
one can obtain supersymmetric QCD, see \cite{BroSuij}. 

The higher order terms 
in the spectral action, which appear with coefficients $f_{-2k} =(-1)^k k!/(2k)! f^{(2k)}(0)$
depending on the derivatives of the test function, and involve higher derivative terms
in the fields, were considered explicitly recently, in work related to renormalization
of the spectral action for gauge theories \cite{WvS1}, \cite{WvS2}, and also in \cite{uncanny}.
In cases where the underlying geometry is very symmetric (space forms) and the Dirac
spectrum is explicitly known, it is also possible to obtain explicit non-perturbative
computations of the spectral action, computed directly as $\Tr(f(D/\Lambda))$, using
Poisson summation formula techniques applied to the Dirac spectrum and its
multiplicities, see \cite{uncanny}, \cite{MaPieTeh}, \cite{MaPieTeh2}, \cite{Teh}.

\subsection{Almost-commutative geometries}

It is also natural to consider a generalization of the product geometry $X\times F$,
where this type of {\em almost-commutative geometry} is generalized to allow for
nontrivial fibrations that are only locally, but not globally, products.  This means
considering almost-commutative geometries that are fibrations over an ordinary
manifold $X$, with fiber a finite noncommutative space $F$. A first instance where
such topologically non-trivial cases were considered in the context of models of
gravity coupled to matter was the Yang--Mills case considered in \cite{BoeSuij}.

In the setting of \cite{BoeSuij}, instead of a product geometry $X\times F$, one considers 
a noncommutative space obtained as an algebra bundle, namely where the
algebra of the full space is isomorphic to sections $\Gamma(X,\cE)$ of
a locally trivial $*$-algebra bundle whose fibers $\cE_x$ are isomorphic to 
a fixed finite-dimensional algebra $\cA_F$. The spectral triple that
replaces the product geometry is then of the form $(C^\infty(X,\cE), L^2(X,\cE\otimes S),
\cD_{\cE})$, where the Dirac operator $\cD_{\cE}=c\circ (\nabla^{\cE}\otimes 1 
+ 1\otimes \nabla^S)$ is defined using the spin connection and a hermitian connection on the
algebra bundle $\cE$ (with respect to an inner product obtained using a faithful
tracial state $\tau_x$ on $\cE_x$).  The spectral triple obtained in this way can
be endowed with a compatible grading and real structure and it is described
in \cite{BoeSuij} in terms of an unbounded Kasparov product of KK-cycles.  In the
Yang--Mills case, where the finite-dimensional algebra is $\cA_F =M_N(\C)$, 
it is then shown in \cite{BoeSuij} that this type of spectral triples describes 
$PSU(N)$-gauge theory with a nontrivial principal bundle and the Yang--Mills
action functional coupled to gravity is recovered from the asymptotic expansion
of the spectral action. 

Even more generally, one can define an almost-commutative geometry with base $X$, as a spectral triple of the form $(\cA,\cH,D)$, where $\cH = L^{2}(X,\cV)$ for $\cV \to X$ a self-adjoint Clifford module bundle, $\cA = C^{\infty}(X,\cE)$ for $\cE \to X$ a unital $*$-algebra sub-bundle of ${\rm End}^+_{\Cl(X)}(\cV)$, and $D$ a symmetric Dirac-type operator on $\cV$; in this context, the compact Riemannian manifold $X$ is no longer required to be spin. A reconstruction theorem for almost commutative geometries (defined in this more general topologically nontrivial sense), was recently obtained in \cite{Cacic2}, as a consequence of the reconstruction theorem for commutative spectral  triples of \cite{Co-reco}. There, the above concrete definition was shown to be equivalent to an abstract definition of almost-commutative geometry with base a commutative unital $*$-algebra, analogous to the abstract definition of commutative spectral triple.

\subsection{Cosmic topology and inflation}

The asymptotic expansion of the spectral action naturally provides 
an action functional for (Euclidean) modified gravity, where in addition
to the ordinary Einstein--Hilbert action with cosmological term one also
has a topological term (the Euler characteristic) and conformal gravity terms
like the Weyl curvature and a conformal coupling of the Higgs field to
gravity.  It also produces the additional bosonic terms: the action for
the Higgs with quartic potential and the Yang--Mills action for the
gauge fields. Thus, it is natural to consider the spectral action as
a candidate action functional for a modified gravity model and study
its consequences for cosmology. 

Cosmological implications of the spectral action, based on the
asymptotic expansion, were considered in \cite{KoMa}, 
\cite{MaPie}, \cite{NeOcSa1}, \cite{NeOcSa2}, \cite{NeSa}, \cite{NeSa2}.
For recent developments in the case of Robertson--Walker metrics see
\cite{ChCo-RW}.

In \cite{MaPieTeh} and \cite{MaPieTeh2} 
the nonperturbative spectral action was computed explicitly for
the 3-dimensional spherical space forms and the flat 3-dimensional
Bieberbach manifolds, via the same type of Poisson summation 
technique first used in \cite{uncanny} for the sphere case. 
The spectral action for Bieberbach manifolds was also computed in \cite{OlSi}. In these
computations one considers the spectral action as a pure gravity
functional (that is, only on the manifold $X$, without the finite geometry $F$).
It is shown that a perturbation $D^2 +\phi^2$ of the Dirac operator 
produces in the nonperturbative spectral action a slow-roll potential $V(\phi)$ 
for the scalar field $\phi$, which can be used as a model for cosmic
inflation. 

While in \cite{uncanny} the computation of the potential $V(\phi)$ is
performed in the model the case of the Higgs field, we do not assume
here (nor in \cite{MaPieTeh}, \cite{MaPieTeh2})  that $\phi$ is necessarily
related to the Higgs field and we only treat $D^2 +\phi^2$ as a scalar
field perturbation of the Dirac operator. In particular, in the
noncommutative geometry models of particle physics the Higgs field
arises because of the presence of a nontrivial noncommutative 
space $F$, and an almost commutative geometry $X\times F$ (the
product of the spacetime manifold $X$ and $F$, or more generally
a nontrivial fibration as analyzed in \cite{Cacic2}). The Higgs field
is described geometrically as the inner fluctuations of the Dirac
operator in the noncommutative direction $F$. On the other hand,
more general fluctuations of the form $D^2 +\phi^2$ are possible
also in the pure manifold case (the pure gravity case), in the absence
of a noncommutative fiber $F$. These are indeed the cases
considered in \cite{MaPieTeh}, \cite{MaPieTeh2}. It is worth 
pointing out that, if one assumes that the field $\phi$ is related
to the Higgs field, then there are very strong constraints coming
from the CMB data, as recently analyzed in \cite{Buck}, which 
make a Higgs-based inflation scenario, as predicted by this
kind of NCG model, incompatible with the measured value of
the top quark mass. However, these constraints do not directly
apply to other scalar perturbations $\phi$, not related to the
Higgs field. 

In this paper we focus on the same topologies considered in
\cite{MaPieTeh}, \cite{Teh} and \cite{MaPieTeh2}. These
include all the most significant candidates for a nontrivial
cosmic topology, widely studied in the theoretical cosmology 
literature (see for example \cite{NiJa}, \cite{Ria1}, \cite{Ria2}).
We refer the reader to \cite{MaPieTeh} and the references
therein, for a more detailed overview of the specific
physical significance of these various topologies. 

In \cite{MaPieTeh}, \cite{Teh}  and \cite{MaPieTeh2} the
nonperturbative spectral action is computed for all the spherical 
space forms $S^3/\Gamma$ and for the flat tori and for all the
flat Bieberbach 3-manifolds (for the latter case see also \cite{OlSi}).
These two classes of manifolds provide a complete classification of all the possible
homogeneous compact 3-manifolds that are either positively curved or flat,
hence they encompass all the possible compact cases of interest to the
problem of cosmic topology (see for instance \cite{Ria1}, \cite{Ria2}). 
It is shown in \cite{MaPieTeh}, \cite{Teh}  and \cite{MaPieTeh2}
that the nonperturbative spectral action for spherical space forms is,
up to an overall constant factor that depends on the order of the
finite group $\Gamma$, the same as that of the sphere $S^3$, hence so
is the slow-roll potential. Similarly, the spectral action and potentials for
the flat Bieberbach manifolds are a multiple of those of the flat torus $T^3$.
In particular, for each such manifold, although the spectra depend explicitly
on the different spin structures, the spectral action does not.  These
results show that, in a model of gravity based on the spectral action
functional, the amplitudes and slow-roll parameters in the power
spectra for the scalar and tensor fluctuation would depend on the
underlying cosmic topology, and hence constraints on these
quantities derived from cosmological data (see \cite{LidLid}, \cite{SmiKa},
\cite{StLy}) may, in principle, be able to distinguish between different
topologies.

Here we discuss a natural question arising from the results of
 \cite{MaPieTeh} and \cite{MaPieTeh2}, namely how the
presence of the finite geometry $F$ may affect the behavior
of the slow-roll inflation potential.  As a setting, we consider here
the case of the spherical space forms $S^3/\Gamma$ as the commutative
base of an almost commutative geometry in the sense of \cite{Cacic2},
where the Clifford module bundle $\cV$ on $S^3/\Gamma$ is the spinor bundle twisted by a flat bundle 
corresponding to a finite-dimensional representation $\alpha: \Gamma \to \GL_N(\C)$ 
of the group $\Gamma$. As the Dirac operator on the almost commutative
geometry we consider the corresponding twisted 
Dirac operator $D^\Gamma_\alpha$ on $S^3/\Gamma$. From the point
of view of the physical model this means that we only focus on the gravity
terms and we do not include the part of the Dirac operator $D_F$ that describes
the matter content and which comes from a finite spectral triple in the fiber direction.

Our main result is that, for any such almost commutative geometry, the
spectral action and the associated slow-roll potential
only differ from those of the sphere $S^3$ by an overall multiplicative
amplitude factor equal to $N/\# \Gamma$. Thus, the only modification
to the amplitude factor in the power spectra is a correction, which
appears uniformly for all topologies, by a multiplicative factor $N$ 
depending on the fiber of the almost commutative geometry.  In terms of
the physical model, this $N$ represents the number of fermions in the theory.
We obtain this by computing the spectral action in its nonperturbative form,
as in \cite{uncanny}, \cite{MaPieTeh}, \cite{MaPieTeh2}, \cite{Teh},
using the Poisson summation formula technique and the explicit
form of the Dirac spectra derived in \cite{twist}.

More generally, one can consider a finite normal Riemannian cover $\Gamma \to \widetilde{M} \to M$ together with a finite-dimensional representation $\alpha : \Gamma \to U(N)$, and compare the spectral action and slow-roll potential on a $\Gamma$-equivariant almost commutative geometry over $\widetilde{M}$, with the spectral action and slow-roll potential on the quotient geometry twisted by $\alpha$, an almost commutative geometry over $M$. Using a nonperturbative heat kernel argument, we find that the spectral action and slow-roll potential over $M$ differ from those over $\widetilde{M}$, up to an error of order $O(\Lambda^{-\infty})$ as $\Lambda \to +\infty$, by an overall multiplicative amplitude equal to $N/\#\Gamma$, recovering both our result above for spherical space forms, as well as the relations obtained in \cite{MaPieTeh}, \cite{MaPieTeh2} and \cite{Teh}. We also obtain the analogous result for the perturbative spectral action.

\subsection{Basic setup}

We recall the basic setting, following the notation of \cite{twist}.  Let $\Gamma \subset SU(2)$
be a finite group acting by isometries on $S^3$, identified with the Lie group $SU(2)$ with
the round metric. The spinor bundle on the spherical form $S^3/\Gamma$ is given by
$S^3\times_\sigma \C^2 \to S^3/\Gamma$, where $\sigma$ is the representation of $\Gamma$
defined by the standard representation of $SU(2)$ on $\C^2$. 

A unitary representation $\alpha: \Gamma \to U(N)$ defines a flat bundle $\cV_\alpha =
S^3 \times_\alpha \C^N$ endowed with a canonical flat connection. By twisting the Dirac
operator with the flat bundle, one obtains an operator $D^\Gamma_\alpha$ on the
spherical form $S^3/\Gamma$ acting on the twisted spinors, that is,  on
the $\Gamma$-equivariant sections $C^\infty(S^3,\C^2\otimes\C^N)^\Gamma$,
where $\Gamma$ acts by isometries on $S^3$ and by $\sigma\otimes\alpha$ on $\C^2\otimes\C^N$;  these are the sections of the twisted spinor bundle $S^3\times_{\sigma\otimes\alpha}
(\C^2\otimes\C^N)\to S^3/\Gamma$. 
Thus, $D^\Gamma_\alpha$ is the restriction of the Dirac operator 
$D\otimes \id_{\C^N}$ to the subspace $C^\infty(S^3,\C^2\otimes\C^N)^\Gamma
\subset C^\infty(S^3,\C^2\otimes\C^N)$. 

\smallskip

This setup gives rise to an almost commutative geometry in the sense of
\cite{Cacic2}, where the twisted Dirac operator $D^\Gamma_\alpha$ 
represents the ``pure gravity" part of the resulting model of gravity coupled 
to matter, while the fiber $\C^N =\cH_F$ determines the fermion content 
of the matter part and can be chosen according to the type of particle 
physics model one wishes to consider (Standard Model with right handed neutrinos,
supersymmetric QCD, for example, as in \cite{ChCoMa}, \cite{BroSuij}, or other
possibilities).  Since we will only be focusing on the gravity terms, we do not need
to specify in full the data of the almost commutative geometry, beyond assigning
the flat bundle $\cV_\alpha$ and the twisted Dirac operator $D^\Gamma_\alpha$,
as the additional data would not enter directly in our computations.

We later generalise this above setup as follows. Let $\Gamma \to \widetilde{M} \to M$ be a finite normal Riemannian covering with $\widetilde{M}$ and $M$ compact, let $\widetilde{\cV} \to \widetilde{M}$ be a $\Gamma$-equivariant self-adjoint Clifford module bundle, and let $\Gamma$-equivariant symmetric Dirac-type operator $\widetilde{D}$ on $\widetilde{\cV}$; let $\alpha : \Gamma \to U(N)$ be a unitary representation. We can then define a self-adjoint Clifford module bundle $\cV_{\alpha}$ on $M$ by $(\widetilde{\cV} \otimes \bC^{N})/\Gamma$, and a symmetric Dirac-type operator $D_{\alpha}$ on $\cV_{\alpha}$ as the image of $\widetilde{D} \otimes 1_{N}$ acting on $\widetilde{\cV} \otimes \bC^{N}$. Again, since we will be focusing on the gravity terms, and since our results will allow for suitable equivariant perturbations of $\widetilde{D}$, we do not need to specify algebra bundles for $\widetilde{\cV}$ or for $\cV_{\alpha}$.

\smallskip

As it is customary when using the spectral action formalism, the
computation in the 4-dimensional case that determines the form
of the inflation potential $V(\phi)$ is obtained in Euclidean signature,
on a compactification along an $S^1_\beta$, of size $\beta$. For a
detailed discussion of this method and of the significance of
the parameter $\beta$, we refer the reader to \S 3.1 of \cite{MaPieTeh2}.

\smallskip

This paper, as well as the previous two parts \cite{MaPieTeh} and \cite{MaPieTeh2},
are aimed at developing the mathematical aspects of these
noncommutative geometry models of cosmology, hence they focus primarily on
computing the non-perturbative form of the spectral action for a relevant class
of spaces, and showing how one can derive from it a slow--roll inflation potential,
and the dependence of this potential from parameters of the model (in the present 
paper, especially the dimension $N$ of the representation). This paper also
focuses on comparing two different methods for computing the nonperturbative
spectral action, respectively based on the Poisson summation formula and on 
heat-kernel techniques. As such, it is addressed primarily to an audience of
mathematicians and mathematical physicists. Ultimately, the success of such
models in cosmology will inevitably depend on the more strictly cosmological
aspects of this project, namely a direct comparison between the model and
the current CMB data. This is indeed the aim of our ongoing collaboration with the
cosmologist Elena Pierpaoli and will be addressed elsewhere, as appropriate.

\section{Poisson summation formula}

Following the method developed in \cite{uncanny} and \cite{MaPieTeh}, \cite{Teh}, we compute the spectral action of the quotient spaces $S^3/\Gamma$ equipped with the twisted Dirac operator corresponding to a finite-dimensional representation $\alpha$ of $\Gamma$ as follows.  We define a finite set of polynomials labeled $P_m^+$, and $P_m^-$ that describe the multiplicities of, respectively, the positive and negative eigenvalues of the twisted Dirac operator. More precisely, the index $m$ takes values in the set of residue classes of the integers modulo $c_{\Gamma}$, where $c_{\Gamma}$ is the exponent of the group $\Gamma$, the least common multiple of the orders of the elements in $\Gamma$. For $k \equiv m  \rm{~mod~} c_{\Gamma}$, $k\geq 1$, and
\begin{equation}
\label{lamb}
\lambda = -1/2 \pm (k+1),
\end{equation}
$P^{\pm}_m(\lambda)$ equals the multiplicity of $\lambda$.

The main technical result we will prove is the following relation between these polynomials:
\begin{equation}
\label{polySums}
\sum_{m=1}^{c_{\Gamma}} P_m^+(u) = \sum_{m=0}^{c_{\Gamma}-1} P_m^-(u) = \frac{N c_{\Gamma}}{\#\Gamma}\left(u^2 - \frac{1}{4} \right).
\end{equation}
Since the polynomial on the right-hand side is a multiple of the polynomial for the
spectral multiplicities of the Dirac spectrum of the sphere $S^3$ (see \cite{uncanny}),
we will obtain from this the relation between the non-perturbative spectral action of
the twisted Dirac operator $D^\Gamma_\alpha$ on $S^3/\Gamma$ and the spectral
action on the sphere, see Theorem \ref{mainthmS3Gamma} below.

Furthermore, we shall show that the polynomials $P_m^+(u)$ match up perfectly with the polynomials $P_m^-(u)$, so that the polynomials $P_m^+(u)$ alone describe the entire spectrum by allowing the parameter $k$ in equation \ref{lamb} to run through all of $\Z$. Namely, what we need to show is that

\begin{equation}
\label{both}
P_m^+(u) = P_{m'}^-(u),
\end{equation}

where for each $m$, $m'$ is the unique number between 0 and $c_{\Gamma} -1$ such that $m + m' +2$ is a multiple of $c_{\Gamma}$, or more precisely,

\begin{equation}
m' = \begin{cases}
c_{\Gamma}-2-m, & \mbox{if } 1 \leq m \leq c_{\Gamma}-2,\\
c_{\Gamma}-1, & \mbox{if } m = c_{\Gamma}-1, \\
c_{\Gamma}-2  & \mbox{if } m = c_{\Gamma}.
\end{cases}
\end{equation}

Define
\begin{equation}
g_m(u) = P^{+}_m(u) f(u/\Lambda).
\end{equation}
Now, we apply the Poisson summation formula to obtain
\begin{align*}
\Tr(f (D/ \Lambda)) &= \sum_{m} \sum_{l \in \Z} g_m(1/2 + c_{\Gamma} l + m + 1)\\
&= \frac{N}{\#\Gamma}\sum_{m} \widehat{g_m}(0) + O(\Lambda^{-\infty})\\
&=  \frac{N}{\#\Gamma}\left(\int_{\R} u^2 f (u/\Lambda) - \frac{1}{4}\int_{\R} f (u/\Lambda) \right) + O(\Lambda^{-\infty}) \\
&= \frac{N}{\#\Gamma}\left(\Lambda^3 \widehat{f}^{(2)}(0) - \frac{1}{4}\Lambda \widehat{f}(0) \right) + O(\Lambda^{-\infty}), \\
\end{align*}
and so we have the main result.

\begin{thm}\label{mainthmS3Gamma}
Let $\Gamma$ be a finite subgroup of $S^3$, and let $\alpha$ be a $N$-dimensional representation of $\Gamma$.  Then the spectral action of $S^3 / \Gamma$ equipped with the twisted Dirac operator is 
\begin{equation}
\Tr f(D / \Lambda) = \frac{N}{\# \Gamma }\left(\Lambda^3 \widehat{f}^{(2)}(0) - \frac{1}{4}\Lambda \widehat{f}(0) \right)+ O(\Lambda^{-\infty}),
\end{equation}
where $\widehat{f}^{(2)}$ denotes the Fourier transform of $u^2 f(u)$.
\end{thm}
Similar computations of the spectral action have also been performed in \cite{MaPieTeh}, \cite{MaPieTeh2}, and \cite{Teh}.  In the sequel we describe how to obtain equation \eqref{polySums},
by explicitly analyzing the cases of the various spherical space forms: lens spaces, dicyclic
group, and binary tetrahedral, octahedral, and icosahedral groups. In all cases we compute
explicitly the polynomials of the spectral multiplicities and check that \eqref{polySums} is
satisfied. Our calculations are based on a result of Cisneros-Molina, \cite{twist}, on the
explicit form of the Dirac spectra of the twisted Dirac operators $D^\Gamma_\alpha$, 
which we recall here below.

\subsection{Twisted Dirac spectra of spherical space forms}
The spectra of the twisted Dirac operators on the quotient spaces are derived in \cite{twist}.  Let us recall the notation and the main results. 

Let $E_k$ denote the $k+1$-dimensional irreducible representation of $SU(2)$ on the space of homogeneous complex polynomials in two variables of degree $k$.
By the Peter--Weyl theorem, one can decompose $C^\infty(S^3,\C)=\oplus_k E_k \otimes E_k^*$ as a sum of irreducible representations of $SU(2)$. This gives that, on 
$C^\infty(S^3,\C^2\otimes\C^N)=\oplus_k E_k \otimes E_k^*\otimes \C^2\otimes\C^N$,
the Dirac operator $D\otimes \id_{\C^N}$ decomposes as $\oplus_k \id_{E_k} \otimes D_k\otimes \id_{\C^N}$, with $D_k: E_k^* \otimes \C^2 \to E_k^* \otimes \C^2$.  Upon
identifying $C^\infty(S^3,\C^2\otimes \C^N)^\Gamma = \oplus_k E_k \otimes \Hom_\Gamma
(E_k, \C^2\otimes\C^N)$, one sees that,  as shown in \cite{twist}, the multiplicities of the 
spectrum of the twisted Dirac operator $D_\alpha^\Gamma$ are given by the 
dimensions $\dim_\C \Hom_\Gamma(E_k, \C^2\otimes\C^N)$, which in turn can
be expressed in terms of the pairing of the characters of the corresponding
$\Gamma$-representation, that is, as 
$\langle \chi_{E_k}, \chi_{\sigma\otimes\alpha} \rangle_\Gamma$. One then obtains
the following:

\begin{thm}\label{mult} {\em (Cisneros-Molina, \cite{twist})}
Let $\alpha :\Gamma \rightarrow GL_N(\C)$ be a representation of $\Gamma$.  Then the eigenvalues of the twisted Dirac operator $D_{\alpha}^{\Gamma}$ on $S^3 / \Gamma$ are
\begin{align*}
-\frac{1}{2}-(k+1)~ &\text{with multiplicity}~ \langle \chi_{E_{k+1}}, \chi_{\alpha} \rangle_{\Gamma} (k + 1), &\text{if $k \geq 0$,} \\
-\frac{1}{2}+(k+1)~ &\text{with multiplicity}~ \langle \chi_{E_{k-1}}, \chi_{\alpha} \rangle_{\Gamma} (k + 1), &\text{if $k \geq 1$.}
\end{align*}
\end{thm}

\begin{prop}\label{recur} {\em (Cisneros-Molina, \cite{twist})}
Let $k = c_{\Gamma} l + m$ with $0 \leq m < c_{\Gamma}$.
\begin{enumerate}
\item If $ -1 \in \Gamma$, then
\[
\langle \chi_{E_k}, \chi_{\alpha} \rangle_{\Gamma} =
	\begin{cases}
		\frac{c_{\Gamma} l}{\# \Gamma } (\chi_{\alpha}(1) + \chi_{\alpha}(-1)) + \langle \chi_{E_m}, \chi_{\alpha} \rangle_{\Gamma}  &\text{if $k$ is even,}  \\
		\frac{c_{\Gamma} l}{\# \Gamma} (\chi_{\alpha}(1) - \chi_{\alpha}(-1)) + \langle \chi_{E_m}, \chi_{\alpha} \rangle_{\Gamma} &\text{if $k$ is odd.}
	\end{cases}
\]
\item If $-1 \notin \Gamma$, then
\[
\langle \chi_{E_k}, \chi_{\alpha} \rangle_{\Gamma} = \frac{N c_{\Gamma}l}{\#\Gamma} + \langle \chi_{E_m}, \chi_{\alpha} \rangle_{\Gamma}.
\]
\end{enumerate}
\end{prop}

\subsection{Lens spaces, odd order}
In this section we consider $\Gamma = \Z _n$, where $n$ is odd.  When $n$ is odd, $-1 \notin \Gamma$, which affects the expression for the character inner products in Proposition \ref{recur}.

For $m \in \{1,\ldots, n\}$, we introduce the polynomials,
\[
P^+_m(u) = \frac{N}{n}u^2 + (\beta _m^{\alpha} -\frac{m N}{n})u + \frac{\beta_m^{\alpha}}{2} - \frac{m N}{2 n} - \frac{N}{4 n},
\]
where
\[
\beta_m^{\alpha} = \langle \chi_{E_{m-1}}, \chi_{\alpha} \rangle_{\Gamma},
\]
and $m$ takes on values in $\{1,2 ,\ldots , n \}$

Using Theorem \ref{mult} and Proposition \ref{recur}, it is easy to see that the polynomials $P^+_m(u)$ describe the spectrum on the positive side of the real line, i.e., $P^+_m(\lambda)$ equals the multiplicity of
\[
\lambda = -1/2 + (k+1), ~ k \geq 1
\]
whenever $k \equiv m  \rm{~mod~}n$.

For the negative eigenvalues, the multiplicities are described by the polynomials
\[
P^-_m(u) = \frac{N}{n}u^2 + \left(\frac{2N}{n} + \frac{mN}{n}- \gamma _m^{\alpha}\right)u + \frac{3N}{4n} + \frac{mN}{2n} - \frac{\gamma _m^{\alpha}}{2},
\]
$m \in \{0 , 1 , \ldots n-1 \}$, i.e., $P^-_m(\lambda)$ equals the multiplicity of the eigenvalue 
\[
\lambda = -1/2 - (k+1), ~ k \geq 0,
\]
whenever $k \equiv m  \rm{~mod~}n$; here $\gamma_m^{\alpha}$ is defined by
\[
\gamma_m^{\alpha} = \langle \chi_{E_{m+1}}, \chi_{\alpha} \rangle_{\Gamma}.
\]

Let us denote the irreducible representations of $\Z _n$ by $\chi_t$, which sends the generator to $\exp (\frac{2 \pi i t}{N})$, for $t$ a residue class of integers modulo $n$.

For the sake of computation, we take $\Z _n$ to be the group generated by
\[
B = 
\left[
\begin{array}{cc}
e^{\frac{2 \pi i}{n}} & 0 \\
0 &  e^{-\frac{2 \pi i}{n}}
\end{array}
\right].
\]
Then in the representation $E_k$, $B$ acts on the basis polynomials $P_j(z_1,z_2)$, $j \in \{0,1, \ldots k \}$ as follows.
\begin{align*}
B \cdot P_j(z_1,z_2) & = P_j\left( (z_1,z_2) B \right) \\
& = P_j(e^{\frac{2\pi i}{n}}z_1,e^{-\frac{2\pi i}{n}}z_2) \\
& = (e^{\frac{2\pi i}{n}}z_1)^{k-j}(e^{-\frac{2\pi i}{n}}z_2)^j \\
& = e^{\frac{2 \pi i}{n}(k-2j)}P_j(z_1,z_2).
\end{align*}
Hence, $B$ is represented by a diagonal matrix with respect to this basis, and we have

\begin{prop}\label{lensChar}
The irreducible characters $\chi_{E_k}$ of the irreducible representations of $SU(2)$ restricted to $\Z_n$, $n$ odd, are decomposed into the irreducible characters $\chi_{[t]}$ of $\Z_n$ by the equation 
\begin{equation}
\chi_{E_k} = \sum_{j=0}^{j = k} \chi_{[k- 2j]}.
\end{equation}
Here, $[t]$ denotes the number from $0$ to $n-1$ to which $t$ is equivalent mod $n$.
\end{prop}

In the case where $-1 \notin \Gamma$, that is to say, when $\Gamma = \Z_n$ where $n$ is odd, by equating coefficients of the quadratic polynomials $P_m^+$ and $P_{m'}^-$, the condition \ref{both} is replaced by one that may be simply checked.

\begin{lem}
\label{bothRepOdd}
Let $\Gamma$ be any finite subgroup of $SU(2)$ such that $-1 \notin \Gamma$. Then condition \ref{both} is equivalent to the condition
\begin{equation}
\beta_m^{\alpha} + \gamma_{m'}^{\alpha} = \begin{cases}
\chi_{\alpha}(1), & \mbox{if } 1 \leq m \leq c_{\Gamma}-2\\
2 \chi_{\alpha}(1), & \mbox{if } m = c_{\Gamma}-1 , c_{\Gamma}
\end{cases},
\end{equation}
where $\alpha$ is an irreducible representation of $\Gamma$.  Furthermore this condition holds in all cases.
\end{lem}

Using Proposition \ref{lensChar}, it is a simple combinatorial matter to see that
\begin{equation}
\label{lensOddPlus}
\sum_{m=1}^{n} \langle \chi_{E_{m-1}}, \chi_{\alpha} \rangle_{\Gamma} = N \frac{n + 1}{2},
\end{equation}
for any representation $\alpha$ of $\Z_ n$

For the argument to go through, one also needs to check the special case
\[
P_{c_{\Gamma}}^+ (1/2) = 0.
\]
By direct evaluation one can check that this indeed holds.

For the negative side, we see that
\begin{equation}
\label{lensOddMinus}
\sum_{m=1}^{n} \langle \chi_{E_{m+1}}, \chi_{\alpha} \rangle_{\Gamma} = N \frac{n + 3}{2},
\end{equation}
for any representation $\alpha$ of $\Z_ n$,
and so

\begin{prop}
\label{lensOdd}
Let $\Gamma$ be cyclic with $\#\Gamma$ odd, and let $\alpha$ be a $N$-dimensional representation of $\Gamma$.  Then
\[
\sum_{m=1}^{n} P_m^+(u) = \sum_{m=0}^{n-1} P_m^-(u) =  Nu^2 - \frac{N}{4}.
\]
\end{prop}

Note that in the statement of Theorem \ref{mult}, the first line holds even if we take $k = -1$, since the multiplicity for this value evaluates to zero.  Therefore, we automatically have 
\[
P_{c_{\Gamma}-1}^- (-1/2) = 0,
\]
which we still needed to check.

\subsection{Lens spaces, even order}

When $n$ is even, we have $-1 \in \Z _n$.  When $ -1 \in \Gamma$, from Theorems \ref{mult} and \ref{recur} it follows that the multiplicity of the eigenvalue
\[
\lambda = 1/2 + l c_{\Gamma} + m,~ l \in \N
\]
is given by, $P_m^+$, $m \in \{1 ,2 ,\ldots, c_{\Gamma} \}$,

\begin{align*}
P_m^+ (u) =  & \frac{1}{\# \Gamma} (\chi_{\alpha}(1) + (-1)^{m+1} \chi_{\alpha}(-1))u^2 + \\
&
\left(\beta_m^{\alpha} -\frac{1}{\#\Gamma}(m(\chi_{\alpha}(1) 
+(-1)^{m+1} \chi_{\alpha}(-1))\right)u  \\
&           + \frac{1}{2} \beta_m^{\alpha} -\frac{1}{4\#\Gamma}(\chi_{\alpha}(1) +(-1)^{m+1} \chi_{\alpha}(-1)) \\
& -  \frac{1}{2\#\Gamma}m(\chi_{\alpha}(1) +(-1)^{m+1} \chi_{\alpha}(-1))).
\end{align*}

The one case that is not clear is that of $\lambda = 1/2$, which is not an eigenvalue of the twisted Dirac operator.  It is not clear from Theorems \ref{mult} and \ref{recur} that 
\begin{equation}
\label{specialCase}
P_{c_{\Gamma}}^+(1/2) = 0,
\end{equation}
and this needs to hold in order for the argument using the Poisson summation formula to go through.  However, by evaluating equation \eqref{specialCase}, we see that one needs to check that
\begin{equation}
\langle \chi_{E_{c_{\Gamma}-1}}, \chi_{\alpha} \rangle = \frac{c_{\Gamma}}{\#\Gamma}(\chi_{\alpha}(1) + (-1)^{c_{\Gamma}+1}\chi_{\alpha}(-1)),
\end{equation}
so that Equation~\ref{specialCase} indeed holds for each subgroup $\Gamma$ and irreducible representation $\alpha$.

\begin{prop}
\label{sumPlus}
For any subgroup $\Gamma \subset S^3$ of even order, the sum of the polynomials $P_m^+$ is
\begin{align*}
\sum_{m = 1}^{c_{\Gamma}}P_m^+ (u)  =  &\frac{c_{\Gamma}}{\#\Gamma}\chi_{\alpha}(1) u^2 \\
&+\left(-\frac{c_{\Gamma}^2 \chi_{\alpha}(1)}{2 \#\Gamma} - \frac{c_{\Gamma}(\chi_{\alpha}(1) - \chi_{\alpha}(-1))}{2\#\Gamma} + \sum_{m=1}^{c_{\Gamma}} \beta_m^{\alpha} \right)u \\
& - \frac{c_{\Gamma}\chi_{\alpha}(1)}{2 \#\Gamma} - \frac{c_{\Gamma}^2 \chi_{\alpha}(1)}{4\#\Gamma} +
\frac{c_{\Gamma}}{4\#\Gamma}\chi_{\alpha}(-1) + \frac{1}{2}\sum_{m=1}^{c_{\Gamma}} \beta_m^{\alpha}
\end{align*}
\end{prop}

Since the coefficients of the polynomial are additive with respect to direct sum, it suffices to consider only irreducible representations.

In the case of lens spaces, $c_{\Gamma} = \#\Gamma$, and $\chi_t(-1) = (-1)^t$.  As a matter of counting, one can see that
\begin{prop}
\label{cyclic}
\[
\sum_{m=1}^{c_{\Gamma}} \beta_m^{t} = 
\begin{cases}
		\frac{n + 2}{2} &\text{if $t$ is even,}  \\
     \frac{n}{2} &\text{if $t$ is odd.}
\end{cases}
\]
\end{prop}

Putting this all into the expression of Proposition \ref{sumPlus}, we have, for an $N$-dimensional representation $\alpha$,
\begin{equation}
\sum_{m = 1}^{c_{\Gamma}}P_m^+ (u) = N\left(u^2 - \frac{1}{4} \right).
\end{equation}

The negative eigenvalues are described by the polynomials
\begin{align*}
P_m^- (u) = &  \frac{1}{\#\Gamma} (\chi_{\alpha}(1) + (-1)^{m+1}\chi_{\alpha}(-1)) u^2 + \\
& \left( \frac{2 + m}{\#\Gamma} (\chi_{\alpha}(1)  + (-1)^{m+1} \chi_{\alpha}(-1)) - \gamma^{\alpha}_m \right) u  \\
&          \frac{3 + 2 m }{4 \# \Gamma}(\chi_{\alpha}(1) + (-1)^{m+1}\chi_{\alpha}(-1)) - \frac{1}{2}\gamma^{\alpha}_m ,
\end{align*}
for $m \in \{0,1, \ldots c_{\Gamma} -1  \}$, so that we have the following proposition.

\begin{prop}
\label{sumMinus}
For any subgroup $\Gamma \subset S^3$ of even order, the sum of the polynomials $P_m^-$ is
\begin{align*}
\sum_{m = 1}^{c_{\Gamma}}P_m^- (u)  =  &\frac{c_{\Gamma}}{\#\Gamma}\chi_{\alpha}(1) u^2 +\\
& \left( \frac{\chi_{\alpha}(1) c_{\Gamma}^2 } { 2 \#\Gamma} + \frac{3\chi_{\alpha}(1) c_{\Gamma}}{2\#\Gamma} + \frac{\chi_{\alpha}(-1) c_{\Gamma}}{2 \#\Gamma } - \sum_{m=0}^{c_{\Gamma}-1} \gamma_m^{\alpha} \right)u~ +\\
&  \frac{\chi_{\alpha}(1)c_{\Gamma}}{2\#\Gamma}+\frac{\chi_{\alpha}(1)c_{\Gamma}^2}{4\#\Gamma}
+\frac{\chi_{\alpha}(-1)c_{\Gamma}}{4\#\Gamma} - \frac{1}{2}\sum_{m=0}^{c_{\Gamma}-1}\gamma_{m}^{\alpha}.
\end{align*}
\end{prop}

By counting, one can see that
\begin{equation}
\sum_{m=0}^{c_{\Gamma}-1} \gamma^t_{m} = 
\begin{cases}
		\frac{n + 4}{2} &\text{if $t$ is even,}\\
    \frac{n+2}{2} &\text{if $t$ is odd.}
\end{cases}
\end{equation}

To complete the computation of the spectral action one still needs to verify the condition \eqref{both}. We have the following lemma, obtained by equating the coefficients of $P_m^+$ and $P_{m'}^-$, that covers the cases of the binary tetrahedral, octahedral and icosahedral groups as well.
\begin{lem}
\label{bothRepEven}
Let $\Gamma$ be any finite subgroup of $SU(2)$ such that $-1 \in \Gamma$ the condition 
\eqref{both} is equivalent to the condition
\begin{equation}
\beta_m^{\alpha} + \gamma_{m'}^{\alpha} = \begin{cases}
\chi_{\alpha}(1)(\chi_{\alpha}(1) + (-1)^{m+1}\chi_{\alpha}(-1)), &\text{if $1 \leq m \leq c_{\Gamma}-2$,}\\
2 \chi_{\alpha}(1)(\chi_{\alpha}(1) + \chi_{\alpha}(-1)), &\text{if $m = c_{\Gamma}-1$,}\\
2 \chi_{\alpha}(1)(\chi_{\alpha}(1) - \chi_{\alpha}(-1)), &\text{if $m = c_{\Gamma}$,}
\end{cases}
\end{equation}
where $\alpha$ is an irreducible representation of $\Gamma$.  Furthermore this condition holds in all cases.
\end{lem}

\subsection{Dicyclic group}

The character table for the dicyclic group of order $4r$ is, for $r$ odd,

\begin{center}
\begin{tabular}{|c|c|c|c|c|c|}
\hline
Class & $1_+$ & $1_-$ & $2_l$ & $r_0$ & $r_1$ \\
\hline
$\psi_t$ & $2$& $2(-1)^t$ & $\zeta_{2r}^{lt} + \zeta_{2r}^{-lt}$ & $0$ & $0$\\
\hline
$\chi_1$ & $1$& $1$ & $1$ & $1$ & $1$\\
\hline
$\chi_2$ & $1$& $-1$ & $(-1)^l$ & $i$ & $-i$\\
\hline
$\chi_3$ & $1$& $1$ & $1$ & $-1$ & $-1$\\
\hline
$\chi_4$ & $1$& $-1$ & $(-1)^l$ & $-i$ & $i$,\\
\hline
\end{tabular}
\end{center}

and for $r$ even,

\begin{center}
\begin{tabular}{|c|c|c|c|c|c|}
\hline
Class & $1_+$ & $1_-$ & $2_l$ & $r_0$ & $r_1$ \\
\hline
$\psi_t$ & $2$& $2(-1)^t$ & $\zeta_{2r}^{lt} + \zeta_{2r}^{-lt}$ & $0$ & $0$\\
\hline
$\chi_1$ & $1$& $1$ & $1$ & $1$ & $1$\\
\hline
$\chi_2$ & $1$& $-1$ & $(-1)^l$ & $i$ & $-i$\\
\hline
$\chi_3$ & $1$& $1$ & $1$ & $-1$ & $-1$\\
\hline
$\chi_4$ & $1$& $-1$ & $(-1)^l$ & $-i$ & $i$\\
\hline
\end{tabular}
\end{center}

Here $\zeta_{2r} = e^{\frac{\pi i}{r}}$, $1 \leq t \leq r-1,$ $1 \leq l \leq r-1$.  The notation for the different conjugacy classes can be understood as follows.  The number indicates the order of the conjugacy class, while a sign in the subscript indicates the sign of the traces of the elements in the conjugacy class as elements of $SU(2)$.

For the dicyclic group of order $4r$, the exponent of the group is 
\[
c_{\Gamma} = 
\begin{cases}
		2r &\text{if $r$ is even,}\\
                   4r & \text{if $r$ is odd.}
\end{cases}
\]

One can decompose the characters $\chi_{E_k}$ into the irreducible characters by inspection, and with some counting obtain the following propositions.
\begin{prop}
\label{dicEven}
Let $\Gamma$ be the dicyclic group of order $4r$, where $r$ is even.
\[
\sum_{m=1}^{c_{\Gamma}} \beta_m^{\alpha} = 
\begin{cases}
	\frac{r}{2} &\text{if $\chi_{\alpha} \in \{\chi_1, \chi_2, \chi_3, \chi_4 \}$,}\\
                   r &\text{if $\chi_{\alpha} = \psi_t$, $t$ is even,}\\
                   r + 1 &\text{if $\chi_{\alpha} = \psi_t$, $t$ is odd;} 	
\end{cases}
\]

\[
\sum_{m=0}^{c_{\Gamma}-1} \gamma_m^{\alpha} = 
\begin{cases}
	\frac{r}{2} +1 &\text{if $\chi_{\alpha} \in \{\chi_1, \chi_2, \chi_3, \chi_4 \}$,}\\
	                   r +2 &\text{if $\chi_{\alpha} = \psi_t$, $t$ is even,} \\
                   r + 1 &\text{if $\chi_{\alpha} = \psi_t$, $t$ is odd.} 	
\end{cases}
\]
\end{prop}

\begin{prop}
\label{dicOdd}
Let $\Gamma$ be the dicyclic group of order $4r$, where $r$ is odd.
\[
\sum_{m=1}^{c_{\Gamma}} \beta_m^{\alpha} = 
\begin{cases}
	2r &\text{if $\chi_{\alpha} \in \{\chi_1,  \chi_3 \}$,}  \\
	2r +1 &\text{ if $\chi_{\alpha} \in \{\chi_2,  \chi_4 \}$,}  \\
         4r &\text{if $\chi_{\alpha} = \psi_t$, $t$ is even,} \\
         4r + 2 &\text{if $\chi_{\alpha} = \psi_t$, $t$ is odd;} 	
\end{cases}
\]

\[
\sum_{m=0}^{c_{\Gamma}-1} \gamma_m^{\alpha} = 
\begin{cases}
	2r + 2&\text{if $\chi_{\alpha} \in \{\chi_1,  \chi_3 \}$,}  \\
	2r +1 &\text{if $\chi_{\alpha} \in \{\chi_2,  \chi_4 \}$,}  \\
         4r + 4 &\text{if $\chi_{\alpha} = \psi_t$, $t$ is even,} \\
         4r + 2 &\text{if $\chi_{\alpha} = \psi_t$, $t$ is odd.} 	
\end{cases}
\]
\end{prop}

\subsection{Binary tetrahedral group}
The binary tetrahedral group has order 24 and exponent 12.
The character table of the binary tetrahedral group is

\begin{center}
\begin{tabular}{|c|c|c|c|c|c|c|c|}
\hline
Class & $1_+$ & $1_-$ & $4_{a+}$ & $4_{b+}$ & $4_{a-}$ & $4_{b-}$ & $6$ \\
\hline
Order & $1$ & $2$ & $6$ & $6$ & $3$ & $3$ & $4$ \\
\hline
$\chi_1$ & $1$& $1$ & $1$ & $1$ & $1$ & $1$ & $1$\\
\hline
$\chi_2$ & $1$& $1$ & $\omega ^2$ & $\omega$ & $\omega$ & $\omega^2$ & $1$\\
\hline 
$\chi_3$ & $1$& $1$ & $\omega$ & $\omega ^2$ & $\omega ^2$ & $\omega$ & $1$\\
\hline
$\chi_4$ & $2$& $-2$ & $1$ & $1$ & $-1$ & $-1$ & $0$\\
\hline
$\chi_5$ & $2$& $-2$ & $\omega ^2$ & $\omega$ & $-\omega$ & $-\omega ^2$ & $0$\\
\hline
$\chi_6$ & $2$& $-2$ & $\omega$ & $\omega^2$ & $-\omega^2$ & $-\omega$ & $0$\\
\hline
$\chi_7$ & $3$& $3$ & $0$ & $0$ & $0$ & $0$ & $-1$\\
\hline
\end{tabular}
\end{center}

Here, $\omega = e^{\frac{2 \pi i}{3}}$. 

For the remaining three groups, we can use matrix algebra to decompose the characters $\chi_{E_k}$.

Let $\chi_j$, $x_j$, $j = 1, 2 \ldots, d$ denote the irreducible characters and representatives of the conjugacy classes of the group $\Gamma$, respectively.  Then, since every character decomposes uniquely into the irreducible ones, we have a unique expression for $\chi_{E_k}$ as the linear combination
\[
\chi_{E_k} = \sum_{j=0}^d c_j^k \chi_j.
\] 
If we let $b= (b_j)$ $j = 1,\ldots, d$ be the column with $b_j = \chi_{E_k}(x_j)$, and let $A = (a_{ij})$ be the $d \times d$ matrix where $a_{ij} = \chi_{j}(x_i)$ and let $c = (c_j^k)$ $j = 1, \ldots d$ be another column,  then 
we have
\[
b = A c;
\]
but $A$ is necessarily invertible by the uniqueness of the coefficient column $c$, and so $c$ is given by
\[
c = A^{-1} b.
\]

By this method, we obtain the following proposition.

\begin{prop}
\label{binTet}
Let $\Gamma$ be the binary tetrahedral group. Then,
\[
\sum_{m=1}^{c_{\Gamma}} \beta_m^{\alpha} = 
\begin{cases}
	3, &\text{if $\chi_{\alpha} \in \{\chi_1, \chi_2, \chi_3 \}$,}  \\
	7, &\text{if $\chi_{\alpha} \in \{\chi_4,  \chi_5, \chi_6 \}$,}  \\
         9, &\text{if $\chi_{\alpha} = \chi_7$;}
\end{cases}
\]

\[
\sum_{m=0}^{c_{\Gamma}-1} \gamma_m^{\alpha} = 
\begin{cases}
	4, &\text{if $\chi_{\alpha} \in \{\chi_1, \chi_2, \chi_3 \}$,}  \\
	7, &\text{if $\chi_{\alpha} \in \{\chi_4,  \chi_5, \chi_6 \}$,}  \\
         12, &\text{if $\chi_{\alpha} = \chi_7$.}
\end{cases}
\]
\end{prop}

\subsection{Binary octahedral group}
The binary octahedral group has order 48 and exponent 24.  The character table of the binary octahedral group is

\begin{center}
\begin{tabular}{|c|c|c|c|c|c|c|c|c|}
\hline
Class & $1_+$ & $1_-$ & $6_+$ & $6_0$ & $6_-$ & $8_+$ & $8_-$ & $12$ \\
\hline
Order & $1$ & $2$ & $8$ & $4$ & $8$ & $6$ & $3$ & $4$ \\
\hline
$\chi_1$ & $1$& $1$ & $1$ & $1$ & $1$ & $1$ & $1$ & $1$ \\
\hline
$\chi_2$ & $1$& $1$ & $-1$ & $1$ & $-1$ & $1$ & $1$ & $-1$ \\
\hline
$\chi_3$ & $2$& $2$ & $0$ & $2$ & $0$ & $-1$ & $-1$ & $0$ \\
\hline
$\chi_4$ & $2$& $-2$ & $\sqrt{2}$ & $0$ & $-\sqrt{2}$ & $1$ & $-1$ & $0$ \\
\hline
$\chi_5$ & $2$& $-2$ & $-\sqrt{2}$ & $0$ & $\sqrt{2}$ & $1$ & $-1$ & $0$ \\
\hline
$\chi_6$ & $3$& $3$ & $-1$ & $-1$ & $-1$ & $0$ & $0$ & $1$ \\
\hline
$\chi_7$ & $3$& $3$ & $1$ & $-1$ & $1$ & $0$ & $0$ & $-1$ \\
\hline
$\chi_8$ & $4$& $-4$ & $0$ & $0$ & $0$ & $-1$ & $1$ & $0$ \\
\hline
\end{tabular}
\end{center}

\begin{prop}
\label{binOct}
Let $\Gamma$ be the binary octahedral group. Then,
\[
\sum_{m=1}^{c_{\Gamma}} \beta_m^{\alpha} = 
\begin{cases}
	6, &\text{if $\chi_{\alpha} \in \{\chi_1, \chi_2 \}$,}  \\
	12, &\text{if $\chi_{\alpha} = \chi_3$,}\\
         13, &\text{if $\chi_{\alpha}  \in \{\chi_4,  \chi_5\}$,} \\
         18, &\text{if $\chi_{\alpha}  \in \{\chi_6, \chi_7 \}$,}\\
         26, &\text{if $\chi_{\alpha} = \chi_8$;} 
\end{cases}
\]

\[
\sum_{m=0}^{c_{\Gamma}-1} \gamma_m^{\alpha} = 
\begin{cases}
	7, &\text{if $\chi_{\alpha} \in \{\chi_1, \chi_2 \}$,}  \\
	14, &\text{if $\chi_{\alpha} = \chi_3$,}\\
         13, &\text{if $\chi_{\alpha}  \in \{\chi_4,  \chi_5\}$,} \\
         21, &\text{if $\chi_{\alpha}  \in \{\chi_6, \chi_7 \}$,}\\
         26, &\text{if $\chi_{\alpha} = \chi_8$.}
\end{cases}
\]
\end{prop}

\subsection{Binary icosahedral group}
The binary icosahedral group has order 120 and exponent 60.
The character table of the binary icosahedral group is

\begin{center}
\begin{tabular}{|c|c|c|c|c|c|c|c|c|c|}
\hline
Class & $1_+$ & $1_-$ & $30$ & $20_+$ & $20_-$ & $12_{a+}$ & $12_{b+}$ & $12_{a-}$ & $12_{b-}$ \\
\hline
Order & $1$ & $2$ & $4$ & $6$ & $3$ & $10$ & $5$ & $5$ & $10$\\
\hline
$\chi_1$ & $1$& $1$ & $1$ & $1$ & $1$ & $1$ & $1$ & $1$ & $1$\\
\hline
$\chi_2$ & $2$& $-2$ & $0$ & $1$ & $-1$ & $\mu$ & $\nu$ & $-\mu$ & $-\nu$\\
\hline
$\chi_3$ & $2$& $-2$ & $0$ & $1$ & $-1$ & $-\nu$ & $-\mu$ & $\nu$ & $\mu$\\
\hline
$\chi_4$ & $3$& $3$ & $-1$ & $0$ & $0$ & $-\nu$ & $\mu$ & $-\nu$ & $\mu$\\
\hline
$\chi_5$ & $3$& $3$ & $-1$ & $0$ & $0$ & $\mu$ & $-\nu$ & $\mu$ & $-\nu$\\
\hline
$\chi_6$ & $4$& $4$ & $0$ & $1$ & $1$ & $-1$ & $-1$ & $-1$ & $-1$\\
\hline
$\chi_7$ & $4$& $-4$ & $0$ & $-1$ & $1$ & $1$ & $-1$ & $-1$ & $1$\\
\hline
$\chi_8$ & $5$& $5$ & $1$ & $-1$ & $-1$ & $0$ & $0$ & $0$ & $0$\\
\hline
$\chi_9$ & $6$& $-6$ & $0$ & $0$ & $0$ & $-1$ & $1$ & $1$ & $-1$\\
\hline
\end{tabular}
\end{center}

Here, $\mu = \frac{\sqrt{5} +1}{2}$, and $\nu = \frac{\sqrt{5}-1}{2}$.

\begin{prop}
\label{binIco}
Let $\Gamma$ be the binary icosahedral group.
\[
\sum_{m=1}^{c_{\Gamma}} \beta_m^{\alpha} = 
\begin{cases}
	15, &\text{if $\chi_{\alpha} =\chi_1$,}  \\
	31, &\text{if $\chi_{\alpha} \in \{\chi_2,  \chi_3 \}$,}  \\
         45, &\text{if $\chi_{\alpha} \in \{\chi_4,  \chi_5 \}$,}  \\
         60, &\text{if $\chi_{\alpha} = \chi_6$,} \\
         62, &\text{if $\chi_{\alpha} = \chi_7$,} \\
         75, &\text{if $\chi_{\alpha} = \chi_8$,} \\
         93, &\text{if $\chi_{\alpha} = \chi_9$;}
\end{cases}
\]
\[
\sum_{m=0}^{c_{\Gamma}-1} \gamma_m^{\alpha} = 
\begin{cases}
	16, &\text{if $\chi_{\alpha} =\chi_1$,}  \\
	31, &\text{if $\chi_{\alpha} \in \{\chi_2,  \chi_3 \}$,}  \\
         48, &\text{if $\chi_{\alpha} \in \{\chi_4,  \chi_5 \}$,}  \\
         64, &\text{if $\chi_{\alpha} = \chi_6$,} \\
         62, &\text{if $\chi_{\alpha} = \chi_7$,} \\
         80, &\text{if $\chi_{\alpha} = \chi_8$,} \\
         93, &\text{if $\chi_{\alpha} = \chi_9$.} 
\end{cases}
\]
\end{prop}

\subsection{Sums of polynomials}
If we input the results of Propositions \ref{cyclic}, \ref{dicEven}, \ref{dicOdd}, \ref{binTet}, \ref{binOct}, \ref{binIco} into Propositions \ref{sumPlus}, \ref{sumMinus}, and also recall Proposition \ref{lensOdd}, we obtain the following.

\begin{prop}
Let $\Gamma$ be any finite subgroup of $SU(2)$ and let $\alpha$ be an $N$-dimensional representation of $\Gamma$.  Then the sums of the polynomials $P_m^+$ and $P_m^-$ are given by
\[
\sum_{m=1}^{c_{\Gamma}}P_m^+ (u) = \sum_{m=0}^{c_{\Gamma}-1}P_m^- (u)  = 
\frac{N c_{\Gamma}}{\#\Gamma}\left( u^2 - \frac{1}{4}\right).
\]
\end{prop}

\section{A heat-kernel argument}

It may at first seem surprising that, in the above calculation, using the
Poisson summation formula and the explicit Dirac spectra, although the
spectra themselves depend in a subtle way upon the representation
theoretic data of the unitary representation $\alpha: \Gamma \to U(N)$,
through the pairing of the characters of representations, the resulting
spectral action only depends upon the dimension $N$ of the representation, the
order of $\Gamma$, and the spectral action on $S^3$.

This phenomenon is parallel to the similar observation in the Poisson
formula computation of the spectral action for the spherical space
forms and the flat Bieberbach manifolds in the untwisted case \cite{MaPieTeh},
\cite{MaPieTeh2}, \cite{Teh}, where one finds that, although the Dirac spectra are different
for different spin structures, the resulting spectral action depends only
on the order $\#\Gamma$ of the finite group and the spectral action on
$S^3$ or $T^3$.

In this section, we give a justification for this phenomenon based on a
heat-kernel computation that recovers the result of Theorem~\ref{mainthmS3Gamma} 
and justifies the presence of the factor $N/\#\Gamma$.

\subsection{Generalities}

We begin with some background on the spectral action for almost commutative spectral triples. In what follows, let $\cL$ denote the Laplace transform, and let $\cS(0,\infty) = \{\phi \in \cS(\R) \mid \phi(x) = 0,\; x \leq 0\}$.

\begin{thm}\label{pthm1}
Let $(A,H,D)$ be a spectral triple of metric dimension $p$, and let $f : \bR \to \bC$ be even. If $|f(x)| = O(|x|^{-\alpha})$ as $x \to \infty$, for some $\alpha > p$, then for any $\Lambda > 0$, $f(D/\Lambda) = f(|D|/\Lambda)$ is trace-class. If, in addition, $f(x) = \cL[\phi](x^{2})$ for some measurable $\phi : \bR_{+} \to \bC$, then
\begin{equation}
	\Tr \left(f(D/\Lambda)\right) =  \int_0^\infty \Tr \left(e^{-s D^2/\Lambda^2}\right) \phi(s) ds.
\end{equation}
\end{thm}

\begin{proof}
Fix $\Lambda>0$. Let $\mu_k$ denote the $k$-th eigenvalue of $D^2$ in increasing order, counted with multiplicity; since $(A,H,D)$ has metric dimension $p$, $(\mu_k+1)^{-p/2} = O(k^{-1})$ as $k \to \infty$, and hence for $k > \dim \ker D$, $\mu_k^{-1} = O(k^{-2/p})$ as $k \to \infty$. By our hypothesis on $f$, then, for $k > \dim \ker D$,
\[
	|f(\mu_k^{1/2}/\Lambda)| = O(k^{-\alpha/p}), \; k \to \infty;
\]
since $2\alpha/p > 1$, this implies that $\sum_{k=1}^\infty f(\mu_k^{1/2}/\Lambda)$ is absolutely convergent, as required.

Now, suppose, in addition, that $f(x) = \cL[\phi](x^{2})$ for some measurable $\phi : [0,\infty) \to \bC$. Then
\begin{align*}
\Tr \left(f(D/\Lambda)\right) &= \sum_{k=1}^\infty \cL[\phi](\mu_k/\Lambda^2)\\
	&= \sum_{k=1}^\infty \int_0^\infty e^{-s \mu_k/\Lambda^2} \phi(s) ds\\
	&= \int_0^\infty \left[ \sum_{k=1}^\infty e^{-s \mu_k/\Lambda^2} \right] \phi(s) ds\\
	&= \int_0^\infty \Tr \left(e^{-s D^2/\Lambda^2}\right) \phi(s) ds,
\end{align*}
as was claimed.
\end{proof}

The above result raises the question of when a function $\phi : \bR_{+} \to \bC$ defines a function $f(x) = \cL[\phi](x^{2})$ such that $f(D/\Lambda)$ is trace-class; a sufficient condition is given by the following lemma.

\begin{lem}
If $\phi \in \cS(0,\infty)$, then $\cL[\phi](s) = O(s^{-k})$ as $s \to +\infty$, for all $k \in \bN$.
\end{lem}

\begin{proof}
Since $\phi \in \cS(0,\infty)$, $\phi^{(k)}$ is a bounded function with $\phi^{(k)}(0) = 0$ for all $k \in \bN$, and hence $s^{n}\cL[\phi](s) = \cL[\phi^{(n)}](s)$ too is bounded, as required.
\end{proof}

As a corollary of the above results, together with the asymptotic expansion of the heat kernel of a generalised Laplacian, we obtain the following basic result on the spectral action for almost commutative spectral triples:

\begin{cor}[cf.~{\cite[Theorem 1]{NVW}}]\label{pcor1}
Let $\cV$ be a self-adjoint Clifford module bundle on a compact oriented Riemannian manifold $M$, and let $D$ be a symmetric Dirac-type operator on $\cV$. Let $f : \bR \to \C$ be of the form $f(x) = \cL[\phi](x^2)$ for $\phi \in \cS(0,\infty)$. Then for $\Lambda > 0$, $f(D/\Lambda)$ is trace-class with
\begin{equation}
	\Tr \left(f(D/\Lambda)\right) = \int_0^\infty \left[ \int_M \tr \left(K(s/\Lambda^2,x,x)\right) d\Vol(x)\right] \phi(s) ds,
\end{equation}
where $K(t,x,y)$ denotes the heat kernel of $D^2$, and $\Tr f(D/\Lambda)$ admits the asymptotic expansion
\begin{equation}
	\Tr \left(f(D/\Lambda)\right) \sim \sum_{k=-\dim M}^{\infty} \Lambda^{-k} \phi_{k} \int_{M}a_{k+\dim M}(x,D^{2}) d\Vol(x), 
\end{equation}
as $\Lambda \to +\infty$, where $a_{n}(x,D^{2})$ is the $n$-th Seeley-DeWitt coefficient of the generalised Laplacian $D^{2}$, and the constants $\phi_{n}$ are given by
\[
	\phi_{n} = \int_0^\infty \phi(s) s^{n/2} ds.
\]
\end{cor}

\begin{proof}
The first part of the claim follows immediately from the fact that
\[
	\Tr(e^{-tD^2}) = \int_M \tr\left(K(t,x,x)\right)d\Vol(x), \quad t>0,
\]
while the second part of the claim follows immediately from the asymptotic expansion
\[
	\tr\left(K(t,x,x)\right) \sim t^{-\dim M/2} \sum_{k = 0}^\infty t^{k/2} a_{k}(x,D^2), \quad t \to 0,
\]
together with the assumption that $\phi$ has rapid decay, so that, in particular, the $\phi_n$ are all finite.
\end{proof}

In fact, since $a_{n}(\cdot,D^{2}) = 0$ for $n$ odd~\cite[Lemma 1.7.4]{Gil}, one has that the asymptotic form of $\Tr \left(f(D/\Lambda)\right)$, as $\Lambda \to +\infty$, is given by
\[ \begin{cases}
		\sum_{n=0}^{\infty} \Lambda^{2(m-n)} \phi_{2(m-n)}  \int_{M}a_{2n}(x,D^{2})d\Vol(x), &\text{if $\dim M = 2m$,}\\
		\sum_{n=0}^{\infty} \Lambda^{2(m-n)+1} \phi_{2(m-n)+1}\int_{M}a_{2n}(x,D^{2})d\Vol(x), &\text{if $\dim M = 2m+1$.}
	\end{cases}
\]
Note also that for $n > 0$,
\[
	\phi_{-n} =  \int_0^\infty \phi(s) s^{-n/2} ds = \frac{1}{\Gamma(n/2)}\int_{0}^{\infty}f(u)u^{n-1}du.
\]

The following result guarantees that the $\phi_k$ can be chosen at will:

\begin{prop}\label{coefficients}
For any $(a_{n}) \in \C^{\Z}$ there exists some $\phi \in \cS(0,\infty)$ such that
\[
	a_{n} = \int_{0}^{\infty} s^{n/2} \phi(s) ds, \quad n \in \Z.
\]
\end{prop}

In fact, this turns out to be a simple consequence of the following result by Dur{\'an} and Estrada, solving the Hamburger moment problem for smooth functions of rapid decay:

\begin{thm}[Dur{\'an}--Estrada {\cite{DuEs}}]\label{DuranEstrada}
For any $(a_{n}) \in \C^{\Z}$ there exists some $\phi \in \cS(0,\infty)$ such that
\[
	a_{n} = \int_{0}^{\infty} s^{n} \phi(s) ds, \quad n \in \Z.
\]
\end{thm}

\begin{proof}[Proof of Proposition~\ref{coefficients}]
By Theorem~\ref{DuranEstrada}, let $\psi \in \cS(0,\infty)$ be such that
\[
	a_{n} = 2\int_{0}^{\infty} s^{n+1} \psi(s) ds, \quad n \in \Z.
\]
Then for $\phi(s) = \psi(\sqrt{s}) \in \cS(0,\infty)$,
\[
	\int_0^\infty s^{n/2}\phi(s)ds = 2\int_0^\infty t^{n+1} \psi(t) dt = a_n, \quad n \in \Z,
\]
as required.
\end{proof}

\subsection{Non-perturbative results}\label{nonperturb}

We now give a non-perturbative heat-kernel-theoretic analysis of the phenomenon mentioned above.

Let $\widetilde{M} \to M$ be a finite normal Riemannian covering with $\widetilde{M}$ and $M$ compact, connected and oriented, and let $\Gamma$ be the deck group of the covering. Let $\widetilde{\cV} \to \widetilde{M}$ be a $\Gamma$-equivariant self-adjoint Clifford module bundle, and let $\widetilde{D}$ be a $\Gamma$-equivariant symmetric Dirac-type operator on $\widetilde{\cV}$. We can therefore form the quotient self-adjoint Clifford module bundle $\cV := \widetilde{\cV}/\Gamma \to M = \widetilde{M}/\Gamma$, with $\widetilde{D}$ descending to a symmetric Dirac-type operator $D$ on $\cV$; under the identification $L^{2}(M,\cV) \cong L^{2}(\widetilde{M},\widetilde{\cV})^{\Gamma}$, we can identify $D$ with the restriction of $\widetilde{D}$ to $C^{\infty}(\widetilde{M},\widetilde{\cV})^{\Gamma}$, where the unitary action $U : \Gamma \to U(L^{2}(\widetilde{M},\widetilde{\cV}))$ is given by $U(\gamma)\xi(\widetilde{x}) := \xi(\widetilde{x}\gamma^{-1})\gamma$.

Our first goal is to prove the following result, relating the spectral action of $D$ to the spectral action of $\widetilde{D}$ in the high energy limit:

\begin{thm}\label{pthm2}
Let $f : \bR \to \bC$ be of the form $f(x) = \cL[\phi](x^{2})$ for $\phi \in \cS(0,\infty)$. Then for $\Lambda > 0$,
\begin{equation}
	\Tr\left(f(D/\Lambda)\right) = \frac{1}{\#\Gamma} \Tr\left(f(\widetilde{D}/\Lambda)\right) + O(\Lambda^{-\infty}), \quad \text{as $\Lambda \to +\infty$.}
\end{equation}
\end{thm}

\begin{rem}
Theorem~\ref{pthm2} continues to hold even when inner fluctuations of the metric are introduced, since for $A \in C^{\infty}(M,\End(\cV))$ symmetric, $D+A$ on $\cV$ lifts to $\widetilde{D} + \widetilde{A}$ on $\widetilde{\cV}$, where $\widetilde{A}$ is the lift of $A$ to $\widetilde{\cV}$.
\end{rem}

To prove this result, we will need a couple of lemmas. First, we have the following well-known general fact:

\begin{lem}\label{plem1}
Let $G$ be a finite group acting unitarily on a Hilbert space $\cH$, and let $A$ be a $G$-equivariant self-adjoint trace-class operator on $H$. Let $\cH^{G}$ denote the subspace of $\cH$ consisting of $G$-invariant vectors. Then the restriction $A\mid \cH^{G}$ of $A$ to $\cH^{G}$ is also trace-class, and
\[
	\Tr\left(A\mid \cH^{G}\right) = \frac{1}{\#G}\sum_{g \in G} \Tr\left(g A\right).
\]
\end{lem}

\begin{proof}
This immediately follows from the observation that $\tfrac{1}{\#G}\sum_{g \in G}g$ is the orthogonal projection onto $\cH^{G}$.
\end{proof}

Now, we can compute the heat kernel trace of $D$ using the heat kernel for $\widetilde{D}$:

\begin{lem}\label{plem2}
For $t > 0$,
\begin{equation}
\begin{array}{rl}
	\displaystyle{\Tr\left(e^{-t D^{2}}\right)} = & \displaystyle{\frac{1}{\#\Gamma}\Tr\left(e^{-t \widetilde{D}^{2}}\right)} \\[3mm]  + & \displaystyle{\frac{1}{\#\Gamma}\sum_{\gamma \in \Gamma \setminus \{e\}} \int_{\widetilde{M}} \tr\left(\rho(\gamma)(\widetilde{x}\gamma^{-1})\widetilde{K}(t,\widetilde{x}\gamma^{-1}, \widetilde{x})\right) d\Vol(\widetilde{x})},
\end{array}	
\end{equation}
where $\widetilde{K}(t,\widetilde{x},\widetilde{y})$ denotes the heat kernel of $\widetilde{D}$, and $\rho$ denotes the right action of $\Gamma$ on the total space $\widetilde{\cV}$.
\end{lem}

\begin{proof}
Let $\gamma \in \Gamma$. Then for any $\xi \in C^{\infty}(\widetilde{M},\widetilde{\cV})$, 
\begin{align*}
	\left(U(\gamma) e^{-t\widetilde{D}^{2}}\right) \xi(\widetilde{x}) 
	&= U(\gamma)\left(\int_{\widetilde{M}} \widetilde{K}(t,\widetilde{x},\widetilde{y})\xi(\widetilde{y})d\Vol(\widetilde{y})\right)\\ 
	&= \rho(\gamma)(\widetilde{x}\gamma^{-1}) \left(\int_{\widetilde{M}} (\widetilde{x}\gamma^{-1})\widetilde{K}(t,\widetilde{x}\gamma^{-1},\widetilde{y}) \xi(\widetilde{y}) d\Vol(\widetilde{y})\right)\\
	&= \int_{\widetilde{M}} \rho(\gamma)(\widetilde{x}\gamma^{-1})\widetilde{K}(t,\widetilde{x}\gamma^{-1},\widetilde{y})\xi(\widetilde{y})d\Vol(\widetilde{y})
\end{align*}
so that the operator $U(\gamma)e^{-t\widetilde{D}^{2}}$ has the integral kernel 
\[
	(t,\widetilde{x},\widetilde{y}) \mapsto \rho(\gamma)(\widetilde{x}\gamma^{-1}) \widetilde{K}(t,\widetilde{x}\gamma^{-1}, \widetilde{y}).
\]
Since $L^{2}(M,\cV) \cong L^{2}(\widetilde{M},\widetilde{\cV})^{\Gamma}$, we can therefore apply Lemma~\ref{plem1} to obtain the desired result.
\end{proof}

Finally, we can proceed with our proof:

\begin{proof}[Proof of Theorem~\ref{pthm2}]
By Corollary~\ref{pcor1} and Lemma~\ref{plem2}, it suffices to show that for $\gamma \in G \setminus \{e\}$,
\[
	\int_{0}^{\infty} \left[ \int_{\widetilde{M}} \tr\left(\rho(\gamma)(\widetilde{x}\gamma^{-1}) \widetilde{K}(s/\Lambda^2,\widetilde{x}\gamma^{-1}, \widetilde{x})\right)d\Vol(\widetilde{x}) \right] \phi(s) ds = O(\Lambda^{-\infty}),
\]
as $\Lambda \to \infty$.

Now, since $\widetilde{M}$ is compact and since the finite group $\Gamma$ acts freely and properly, 
\[
	\inf_{(\widetilde{x},\gamma) \in \widetilde{M} \times \Gamma} d(\widetilde{x}\gamma^{-1},\widetilde{x}) = \min_{(\widetilde{x},\gamma) \in \widetilde{M} \times \Gamma} d(\widetilde{x}\gamma^{-1},\widetilde{x}) > 0.
\]
Hence, by~\cite[Proposition 3.24]{Kah}, there exist constants $C>0$, $c>0$ such that
\[
	\sup_{\widetilde{x}\in\widetilde{M}}\|\widetilde{K}(t,\widetilde{x}\gamma^{-1},\widetilde{x})\|_{2} \leq Ce^{-c/t}, \quad t > 0,
\]
for $\|\cdot\|_{2}$ the fibre-wise Hilbert-Schmidt norm, implying, in turn, that for every $n \in \bN$ there exists a constant $C_{n} > 0$ such that
\[
	\sup_{\widetilde{x}\in\widetilde{M}}\|\widetilde{K}(t,\widetilde{x}\gamma^{-1},\widetilde{x})\|_{2} \leq C_{n}t^{n}, \quad t > 0.
\]
Hence, for each $n \in \bN$,
\begin{align*}
	&\left| \int_{0}^{\infty} \left[ \int_{\widetilde{M}} \tr\left(\rho(\gamma)(\widetilde{x}\gamma^{-1}) \widetilde{K}(s/\Lambda^2,\widetilde{x}\gamma^{-1}, \widetilde{x})\right)d\Vol(\widetilde{x}) \right] \phi(s) ds \right|	\\
	\leq \; &\int_{0}^{\infty} \Vol(M)  \left(\sup_{\widetilde{x}\in\widetilde{M}} \|\rho(\gamma)(\widetilde{x})\|_{2}\right) \left( \sup_{\widetilde{x} \in \widetilde{M}}\|\widetilde{K}(s/\Lambda^2,\widetilde{x}\gamma^{-1},\widetilde{x})\|_{2}\right) |\phi(s)| ds\\
	\leq \; &\Vol(M) \cdot \left(\sup_{\widetilde{x}\in\widetilde{M}} \|\rho(\gamma)(\widetilde{x})\|_{2}\right) \cdot C_{n} \int_{0}^{\infty} (s/\Lambda^{2})^{n}|\phi(s)|ds\\
	= \; & \left( \Vol(M) \cdot \left(\sup_{\widetilde{x}\in\widetilde{M}} \|\rho(\gamma)(\widetilde{x})\|_{2}\right) \cdot C_{n} \cdot \int_{0}^{\infty}s^{n}|\phi(s)|ds\right) \Lambda^{-2n},
\end{align*}
yielding the desired result.
\end{proof}

Now, let $\alpha : \Gamma \to \GL_{N}(\C)$ be a representation of $\Gamma$; by endowing $\bC^{N}$ with a $\Gamma$-equivariant inner product, we take $\alpha : \Gamma \to U(N)$. Since $\widetilde{M} \to M$ is a principal $\Gamma$-bundle, we form the associated Hermitian vector bundle $\cF := \widetilde{M} \times_{\alpha} \bC^{N} \to M$; since $\Gamma$ is finite, we endow $\cF$ with the trivial flat connection $d$. We can therefore form the self-adjoint Clifford module bundle $\cV \otimes \cF \to M$, which admits the symmetric Dirac-type operator $D_{\alpha}$ obtained from $D$ by twisting by $d$, that is,
\[
	D_{\alpha} = D \otimes 1 + c(1 \otimes d),
\]
where $c$ denotes the Clifford action on $\cV \otimes \cF$.

We now obtain the following generalisation of Theorem~\ref{pthm2}, which explains the factor of $N/\#\Gamma$ appearing in Theorem~\ref{mainthmS3Gamma} above:

\begin{thm}\label{pthm3}
Let $f  : \bR \to \bC$ be of the form $f(x) = \cL[\phi](x^{2})$ for $\phi \in \cS(0,\infty)$. Then for $\Lambda > 0$,
\begin{equation}
	\Tr\left(f(D_{\alpha}/\Lambda)\right) = \frac{N}{\#\Gamma} \Tr\left(f(\widetilde{D}/\Lambda)\right) + O(\Lambda^{-\infty}), \quad \text{as $\Lambda \to +\infty$.}
\end{equation}
\end{thm}

\begin{rem}
This result is again compatible with inner fluctuations of the metric, insofar as if $A \in C^{\infty}(M,\End(\cV))$ is symmetric, then $D_{\alpha} + A \otimes 1$ on $\cV \otimes \cF$ is induced from $\widetilde{D} + \widetilde{A}$ on $\widetilde{\cV}$, where $\widetilde{A}$ is $A$ viewed as a $\Gamma$-equivariant element of $C^{\infty}(\widetilde{M},\End(\widetilde{\cV}))$.
\end{rem}

\begin{proof}[Proof of Theorem~\ref{pthm3}]
On the one hand, consider the trivial bundle $\widetilde{\cF} := \widetilde{M} \times \bC^{N}$ over $\widetilde{M}$, together with the trivial flat connection $d$. Then for the action $(\widetilde{x},v)\gamma := (\widetilde{x}\gamma,\alpha(\gamma^{-1})v)$, $\widetilde{\cF}$ is a $\Gamma$-equivariant Hermitian vector bundle, and $d$ is a $\Gamma$-equivariant Hermitian connection on $\widetilde{\cF}$. Then, by taking the tensor product of $\Gamma$-actions, we can endow $\widetilde{\cV}\otimes\widetilde{\cF}$ with the structure of a $\Gamma$-equivariant self-adjoint Clifford module bundle, admitting the $\Gamma$-equivariant symmetric Dirac-type operator $\widetilde{D_{\alpha}} = \widetilde{D} \otimes 1 + c(1 \otimes d)$. As a vector bundle, however, we may simply identify $\widetilde{\cV} \otimes \widetilde{\cF}$ with $\widetilde{\cF}^{\oplus N}$, in which case we may identify $\widetilde{D_{\alpha}}$ with $\widetilde{D} \otimes 1_{N}$.

On the other hand, by construction, the bundle $\cF$ defined above is the quotient of $\widetilde{\cF}$ by the action of $\Gamma$. Hence, under the action of $\Gamma$, the quotient of $\widetilde{\cV} \otimes \widetilde{\cF}$ is the the self-adjoint Clifford module bundle $\cV \otimes \cF$, with $\widetilde{D_{\alpha}}$ descending to the operator $D \otimes 1 + c(1 \otimes d) = D_{\alpha}$.

Finally, by Theorem~\ref{pthm2} and our observations above,
\begin{align*}
	\Tr\left(f(D_{\alpha}/\Lambda)\right) 
	&= \frac{1}{\#\Gamma}\Tr\left(f(\widetilde{D_{\alpha}}/\Lambda)\right) + O(\Lambda^{-\infty})\\
	&= \frac{1}{\#\Gamma}\Tr\left(f(\widetilde{D}/\Lambda) \otimes 1_{N}\right) + O(\Lambda^{-\infty})\\
	&= \frac{N}{\#\Gamma}\Tr\left(f(\widetilde{D}/\Lambda)\right) + O(\Lambda^{-\infty}), \quad &\text{as $\Lambda \to +\infty$,}
\end{align*}
as was claimed.
\end{proof}

One can apply these results to give a quick second proof of Theorem~\ref{mainthmS3Gamma}. 

\begin{proof}[Second proof of Theorem~\ref{mainthmS3Gamma}]
Recall that $\Gamma \subset SU(2)$ is a finite group acting by isometries on $S^{3}$, identified with $SU(2)$ endowed with the round metric, and that $\alpha : \Gamma \to U(N)$ is a representation. Since $S^{3}$ is parallelizable and $\Gamma$ acts by isometries, the spinor bundle $\C^2 \to \cS_{S^{3}} \to S^{3}$ and the Dirac operator $\Dirac_{S^{3}}$ are trivially $\Gamma$-equivariant. Then, by construction, the Dirac-type operator $D_{\alpha}^{\Gamma}$ on $\cS_{S^3} \otimes \cV_{\alpha}$ is precisely the induced operator $D_{\alpha}$ corresponding to $\widetilde{D} = \Dirac_{S^{3}}$, so that by Theorem~\ref{pthm3},
\[
	\Tr\left(f(D_{\alpha}/\Lambda)\right) = \frac{N}{\#\Gamma}\Tr\left(f(\Dirac_{S^{3}}/\Lambda)\right) + O(\Lambda^{-\infty}), \quad \text{as $\Lambda \to +\infty$.}
\]
However, by~\cite[\S 2.2]{uncanny}, one has that
\[
	\Tr\left(f(\Dirac_{S^{3}}/\Lambda)\right) = \Lambda^{3}\widehat{f^{(2)}}(0)-\frac{1}{4}\Lambda\widehat{f}(0) + O(\Lambda^{-\infty}),
\]
where $\widehat{f^{(2)}}$ denotes the Fourier transform of $u^{2}f(u)$. Hence,
\[
	\Tr\left(f(D_{\alpha}/\Lambda)\right) = \frac{N}{\#\Gamma}\left(\Lambda^{3}\widehat{f^{(2)}}(0)-\frac{1}{4}\Lambda\widehat{f}(0)\right) + O(\Lambda^{-\infty}),
\]
as required.
\end{proof}

\subsection{Perturbative results}

Let us now turn to the perturbative picture. In light of Corollary~\ref{pcor1}, it suffices to compare the Seeley-DeWitt coefficients of $\widetilde{D}^{2}$ with those of $D^{2}$ and $D_{\alpha}^{2}$.

\begin{prop}\label{pprop1}
Let $\widetilde{D}$ and $D$ be as above. Let $\pi: \widetilde{M} \to M$ denote the quotient map. Then for all $n \in \bN$,
\[
	a_{n}(\pi(\widetilde{x}),D^{2}) = a_{n}(\widetilde{x},\widetilde{D}^{2}), \quad \widetilde{x} \in \widetilde{M},
\]
and hence
\[
	\int_M a_n(x,D^2)d\Vol(x) = \frac{1}{\#\Gamma}\int_{\widetilde{M}} a_n(\widetilde{x},\widetilde{D}^2) d\Vol(\widetilde{x}).
\]
\end{prop}

\begin{proof}
By~\cite[Lemma 4.8.1]{Gil}, there exist a unique connection $\nabla$ and endomorphism $E$ on $\cV$ such that $D^{2} = \nabla^{\ast}\nabla - E$, and similarly a unique connection $\widetilde{\nabla}$ and endomorphism $\widetilde{E}$ on $\widetilde{\cV}$ such that $\widetilde{D}^{2} = \widetilde{\nabla}^{\ast}\widetilde{\nabla}$. Since $\widetilde{D}^{2}$ is the lift of $D^{2}$ to $\widetilde{\cV}$, it follows by uniqueness that $\widetilde{\nabla}$ and $\widetilde{E}$ are the lifts of $\nabla$ and $E$, respectively, to $\widetilde{\cV}$ as well.

Now, since the finite group $\Gamma$ acts freely and properly on $\widetilde{M}$, let $\{(U_{\alpha},\Psi_{\alpha})\}$ be an atlas for $\widetilde{M}$ such that for each $\alpha$, $\pi\mid U_{\alpha} : U_{\alpha} \to \pi(U_{\alpha})$ is an isometry. Hence, the local data defining $a_{n}(\cdot,D^{2})$ on $U_{\alpha}$ lifts to the local data defining $a_{n}(\cdot,\widetilde{D}^{2})$ on $\pi^{-1}(U_{\alpha})$; since for $P = \nabla^\ast\nabla - E$ a generalised Laplacian on a Hermitian vector bundle $\cE \to X$, each $a_n(\cdot,P)$ is given by a universal polynomial in the Riemannian curvature of $X$, the curvature of $\nabla$, $E$, and their respective covariant derivatives~\cite[\S 4.8]{Gil}, it therefore follows that $a_{n}(\cdot,\widetilde{D}^{2})$ is indeed the lift to $\widetilde{M}$ of $a_{n}(\cdot,D^{2})$, as was claimed.
\end{proof}

\begin{prop}\label{pprop2}
Let $\widetilde{D}$ and $D_{\alpha}$ be as above. Then for all $n \in \bN$,
\[
	a_n(\pi(\widetilde{x}),D_\alpha^2) = N a_n(\widetilde{x},\widetilde{D}^2), \quad \widetilde{x} \in \widetilde{M},
\]
and hence
\[
	\int_M a_{n}(x,D_{\alpha}^{2})d\Vol(x) = \frac{N}{\#\Gamma}\int_M a_{n}(\widetilde{x},\widetilde{D}^{2})d\Vol(\widetilde{x}).
\]
\end{prop}

\begin{proof}
On the one hand, by~\cite[Lemma 1.7.5]{Gil}, $a_{n}(\cdot,\widetilde{D}^{2} \otimes 1_{N}) = N a_{n}(\cdot,\widetilde{D}^{2})$, for $\widetilde{D}^{2} \otimes 1_{N}$ on $\widetilde{\cV}^{\oplus N}$. On the other hand, $\widetilde{D}^{2}\otimes 1_{N}$ is the lift to $\widetilde{\cV}^{\oplus N}$ of $D_{\alpha}^{2}$ on $\cV \otimes \cF$, so that by Proposition~\ref{pprop1}, $a_{n}(\cdot,\widetilde{D}^{2}\otimes 1_{N})$ is the lift to $\widetilde{M}$ of $a_{n}(\cdot,D_{\alpha}^{2})$. Hence, $N a_{n}(\cdot,\widetilde{D}^{2})$ is the lift to $\widetilde{M}$ of $a_{n}(\cdot,D_{\alpha}^{2})$, as required.
\end{proof}

Let us now apply these results to the Dirac operator $\Dirac_{S^{3}}$ on the round $3$-sphere $S^{3}$, together with a finite subgroup $\Gamma$ of $SU(2)$ acting freely and properly on $S^{3} \cong SU(2)$, and a representation $\alpha : \Gamma \to U(N)$. Since $\Dirac_{S^{3}}^{2} = (\nabla^{S})^{\ast}\nabla^{S} + \tfrac{3}{2}$ by the Lichnerowicz formula, it follows from~\cite[Theorem 4.8.16]{Gil} that
\begin{align*}
	\int_{S^{3}}a_{0}(x,\Dirac_{S^{3}}^{2})d\Vol(x) &= \int_{S^{3}}(4\pi)^{-3/2}\tr(\id)d\Vol(x) = \frac{\sqrt{\pi}}{2},\\
	\int_{S^{3}}a_{2}(x,\Dirac_{S^{3}}^{2})d\Vol(x) &= \int_{S^{3}} (4\pi)^{-3/2}\tr\left(\frac{6}{6}\id - \frac{3}{2}\id\right)d\Vol(x) = -\frac{\sqrt{\pi}}{4}.
\end{align*}
Since the operator $D_{\alpha}^{\Gamma}$ is precisely $D_{\alpha}$ as induced by $\widetilde{D} = \Dirac_{S^{3}}$, it therefore follows by Proposition~\ref{pprop2} that
\begin{align*}
	\int_{S^{3}/\Gamma}a_{0}(y,(D_{\alpha}^{\Gamma})^{2})d\Vol(y) &= \frac{N}{\#\Gamma}\int_{S^{3}}a_{0}(x,\Dirac_{S^{3}}^{2}) d\Vol(x) = \frac{N\sqrt{\pi}}{(\#\Gamma)2},\\
	\int_{S^{3}/\Gamma}a_{2}(y,(D_{\alpha}^{\Gamma})^{2})d\Vol(y) &= \frac{N}{\#\Gamma}\int_{S^{3}}a_{2}(x,\Dirac_{S^{3}}^{2}) d\Vol(x) = -\frac{N\sqrt{\pi}}{(\#\Gamma)4}.\\
\end{align*}
Finally, one has that
\begin{align*}
	\phi_{-3} &= \frac{2}{\Gamma(3/2)}\int_{0}^{\infty}f(u)u^{2}du = \frac{2}{\sqrt{\pi}} \int_{-\infty}^{\infty}f(u)u^{2}du = \frac{2}{\sqrt{\pi}}\widehat{f^{(2)}}(0),\\
	\phi_{-1} &= \frac{2}{\Gamma(1/2)}\int_{0}^{\infty}f(u)du = \frac{1}{\sqrt{\pi}} \int_{-\infty}^{\infty}f(u)du = \frac{1}{\sqrt{\pi}}\widehat{f}(0),
\end{align*}
where $\widehat{f^{(2)}}$ is the Fourier transform of $f(u)u^{2}$. Hence,
\begin{align*}
	\Tr\left(f(\Dirac_{S^{3}}/\Lambda)\right) &\sim \Lambda^{3} \phi_{-3} \int_{S^{3}} a_{0}(x,\Dirac_{S^{3}}^{2}) d\Vol(x) \\  & +  \Lambda\phi_{-1}\int_{S^{3}}a_{2}(x,\Dirac_{S^{3}}^{2})d\Vol(x) + O(\Lambda^{-1})\\
	&= \Lambda^{3}\widehat{f^{(2)}}(0) - \frac{1}{4}\Lambda\widehat{f}(0) + O(\Lambda^{-1}),
\end{align*}
and
\begin{align*}
	\Tr\left(f(D_{\alpha}^{\Gamma})\right) &\sim \Lambda^{3} \phi_{-3}\int_{S^{3}/\Gamma}a_{0}(y,(D_{\alpha}^{\Gamma})^{2})d\Vol(y) \\ & + \Lambda\phi_{-1}\int_{S^{3}/\Gamma}a_{2}(y,(D_{\alpha}^{\Gamma})^{2})d\Vol(y) + O(\Lambda^{-1})\\
	&= \frac{N}{\#\Gamma}\left(\Lambda^{3}\widehat{f^{(2)}}(0) - \frac{1}{4}\Lambda\widehat{f}(0)\right) + O(\Lambda^{-1}),
\end{align*}
which is indeed consistent with Theorem~\ref{mainthmS3Gamma}.

\section{The inflation potential and the power spectra}

It was shown in \cite{MaPieTeh}, \cite{MaPieTeh2} that for a 3-manifold $Y$ that is
a spherical space form $S^3/\Gamma$ or a flat Bieberbach manifold (a quotient of
the flat torus $T^3$ by a finite group action), the non-perturbative spectral action
determines a slow-roll potential for a scalar field $\phi$ by setting
$$ \Tr(h((D_{Y\times S^1}^2+\phi^2)/\Lambda^2)) - \Tr(h(D^2_{Y\times S^1}/\Lambda^2))=
V_Y(\phi), $$
up to terms of order $O(\Lambda^{-\infty})$, where, in the spherical space form case
the potential is of the form 
$$ V_Y(\phi) = \pi \Lambda^4 \beta a^3 \cV_Y(\frac{\phi^2}{\Lambda^2}) + \frac{\pi}{2}\Lambda^2
\beta a \cW_Y(\frac{\phi^2}{\Lambda^2}), $$
for $h$ the test function for the computation
of the spectral action on the 4-manifold $Y\times S^1$, 
$a>0$ the radius of the sphere and $\beta>0$ the size of the circle compactification
$S^1$.  The functions $\cV_Y$ and $\cW_Y$ are of the form
\begin{equation}\label{VWYS3}
\cV_Y(x) = \lambda_Y \, \cV_{S^3}(x) \ \  \text{ and } \ \  \cW_Y(x) = \lambda_Y \, \cW_{S^3}(x),
\end{equation}
where, for $Y = S^3 /\Gamma$, the factor $\lambda_Y=(\# \Gamma)^{-1}$, and
\begin{equation}\label{VWS3}
\cV_{S^3}(x) = \int_0^\infty u \, (h(u+x) - h(u))\, du  \ \ \  \text{ and } \ \ \ 
\cW_{S^3}(x) = \int_0^x h(u)\, du.
\end{equation}
Thus, the potential satisfies 
\begin{equation}\label{VYS3}
V_Y(\phi) = \lambda_Y \, V_{S^3}(\phi) = \frac{V_{S^3}(\phi)}{\# \Gamma}.
\end{equation}

The slow-roll potential $V_Y(\phi)$ can be used as a model for cosmological inflation.
As such, it determines the behavior of the power spectra $\cP_{s,Y}(k)$ and $\cP_{t,Y}(k)$
for the density fluctuations and the gravitational waves, respectively given in the form
\begin{equation}\label{PstV}
\cP_s(k) \sim \frac{1}{M_{Pl}^6} \frac{V^3}{(V^\prime)^2} \ \  \text{ and } \ \ 
\cP_t(k) \sim \frac{V}{M_{Pl}^4},
\end{equation}
with $M_{Pl}$ the Planck mass, see \cite{SmiKa} and \cite{MaPieTeh2}
for more details. Including second order terms, these can be written also
as power laws as in \cite{SmiKa}, 
\begin{equation}\label{powerlawP}
\begin{array}{rl}
\cP_s(k) \sim & \cP_s(k_0) 
\displaystyle{\left(\frac{k}{k_0} \right)^{1 - n_s + \frac{\alpha_s}{2} \log(k/k_0)}}
\\[3mm]  \cP_t(k) \sim & \cP_t(k_0) \displaystyle{\left(\frac{k}{k_0} 
\right)^{n_t + \frac{\alpha_t}{2} \log(k/k_0)}} , \end{array}
\end{equation}
where the exponents 
also depend on the slow roll potentials through certain slow-roll parameters. 
Since, as already
observed in \cite{MaPieTeh}, \cite{MaPieTeh2}, the slow-roll parameters are not
sensitive to an overall multiplicative scaling factor in the potential, 
we focus here only on the
amplitude only, which, as shown in \cite{MaPieTeh2}, correspondingly
changes by a multiplicative factor. Namely, in the case of a spherical
space form with the spectral action computed for the untwisted Dirac
operator, one has
\begin{equation}\label{powerlawPY1}
\begin{array}{rl}
\cP_{s,Y}(k) \sim & \lambda_Y \, \cP_s(k_0) 
\displaystyle{\left(\frac{k}{k_0} \right)^{1 - n_{s,S^3} + \frac{\alpha_{s,S^3}}{2} \log(k/k_0)}}
\\[3mm]  \cP_{t,Y}(k) \sim & \lambda_Y \, \cP_t(k_0) \displaystyle{\left(\frac{k}{k_0} 
\right)^{n_{t,S^3} + \frac{\alpha_{t,S^3}}{2} \log(k/k_0)}} , \end{array}
\end{equation}
where, as above, $\lambda_Y = 1/\#\Gamma$.

The amplitude and the exponents of the power law are parameters 
subject to constraints coming from
cosmological observational data, as discussed in \cite{LidLid}, \cite{SmiKa}, 
\cite{StLy}, so that, in principle, such data may be able to constrain the
possible cosmic topologies in a model of gravity based on the spectral action.
To this purpose, it is important to understand how much the amplitude
and the slow-roll parameter are determined by the model. A discussion of
the role of the parameters $\Lambda$, $a$, and $\beta$ is included in
\cite{MaPieTeh2}, while here we focus on how the coupling of gravity
to matter affects these parameters. 

By directly comparing the argument given in \cite{MaPieTeh} proving
\eqref{powerlawPY1} with the result of Theorem~\ref{mainthmS3Gamma} 
above, we see that, in our case, we obtain then the following 
version of \eqref{powerlawPY1}, modified by an overall multiplicative
factor $N$, the total number of fermions in the model of gravity
coupled to matter. 

\begin{prop}\label{amplNGamma}
For a spherical space form $Y=S^3/\Gamma$,  consider
the slow-roll potential $V_{Y,\alpha}(\phi)$ determined by the nonperturbative 
spectral action
$$ \Tr(h((D_{\alpha, Y \times S^1}^2+\phi^2)/\Lambda^2)) - \Tr(h(D^2_{\alpha,Y\times S^1}/\Lambda^2)) = V_{Y,\alpha}(\phi), $$
where $D_{\alpha, Y\times S^1}$ is the Dirac operator induces on the product
geometry $Y\times S^1$ by the twisted Dirac operator $D_\alpha^\Gamma$ on $Y$.
Then the associated power spectra as in \eqref{PstV}, \eqref{powerlawP} satisfy
\eqref{powerlawPY1}, with $\lambda_Y =N/ \#\Gamma$.
\end{prop}

\subsection{Inflation potential in the heat kernel approach}
Let us now consider inflation potentials on space-times of the form $M \times S^1_\beta$ for $M$ compact oriented Riemannian and odd-dimensional, arising from general almost commutative triples over $M$.

Let $D$ is a symmetric Dirac-type operator on a self-adjoint Clifford module bundle $\cV \to M$, with $M$ compact oriented Riemannian and odd-dimensional, and let $\Dirac_\beta$ be the Dirac operator with simple spectrum $\tfrac{1}{\beta}(\Z + \tfrac{1}{2})$ on the trivial spinor bundle $\C \to \cS_{S^1_\beta} \to S^1_\beta$. We may immediately generalise the construction of~\cite[\S 2.3]{uncanny} to obtain an odd symmetric Dirac-type operator $D_{M \times S^1_\beta}$ on the self-adjoint Clifford module bundle $\left(\cV \boxtimes \cS_{S^1_\beta}\right)^{\oplus 2} \to M \times S^1_\beta$. Hence, we may define an inflation potential $V_M : C^\infty\left(M \times S^1_\beta\right) \to \bR$ by 
\[
	V_M(\phi) := \Tr\left(h((D_{M \times S^1_\beta}^2+\phi^2)/\Lambda^2)\right) - \Tr\left(h(D_{M \times S^1_\beta}^2/\Lambda^2)\right),
\]
where $h = \cL[\psi]$ for $\psi \in \cS(0,\infty)$; note that $D_{M \times S^1_\beta}^2 + \phi^2$ has heat trace
\[
	\Tr\left(e^{-t \left(D_{M \times S^1_\beta}^2 +\phi^2\right)}\right) = 2\Tr\left(e^{-t \Dirac_\beta^2}\right)\Tr\left(e^{-tD^2}\right)e^{-\phi^2 t}
\]
for $\phi$ locally constant.

Let $\Gamma \to \widetilde{M} \to M$, $\widetilde{\cV} \to \widetilde{M}$, $\cV \to M$, $\widetilde{D}$, $D$, $\alpha$, $\cF \to M$ and $D_\alpha$ be defined as in Subsection~\ref{nonperturb}, with $M$ and $\widetilde{M}$ odd-dimensional, generalising the discussion above of $\Gamma \to S^3 \to Y$. On the one hand, we may form odd Dirac-type operators $\widetilde{D}_{\tilde{M} \times S^1_\beta}$, $D_{M \times S^1_\beta}$, and $D_{\alpha, M \times S^1_\beta}$ from $\widetilde{D}$, $D$ and $D_\alpha$, respectively, as above. On the other hand, if one trivially extends the action of $\Gamma$ on $\widetilde{M}$ to $\widetilde{M} \times S^1_\beta$ and the action on $\widetilde{\cV} \to \widetilde{M}$ to $\left(\cV \boxtimes \cS_{S^1_\beta}\right)^{\oplus 2} \to M \times S^1_\beta$, then $D_{\widetilde{M} \times S^1_\beta}$ becomes a $\Gamma$-equivariant Dirac-type operator on $\left(\cV \boxtimes \cS_{S^1_\beta}\right)^{\oplus 2}$, and the constructions of Subsection~\ref{nonpertu
 rb} applied to the $\Gamma$-equivariant Dirac-type operator $D_{\widetilde{M} \times S^1_\beta}$ reproduce precisely the Dirac-type operators  $D_{M \times S^1_\beta}$, and $D_{\alpha, M \times S^1_\beta}$.

Now, let $V_{\widetilde{M}}$, $V_M$, and $V_{M,\alpha}$ denote the inflation potentials corresponding to $\tilde{D}$, $D$, and $D_\alpha$, respectively, which we all view as nonlinear functionals on $C^\infty(\widetilde{M} \times S^1_\beta,\bR)^\Gamma \cong C^\infty(M \times S^1_\beta,\bR)$. Then, since we also have that $D_{\widetilde{M}\times S^1_\beta}^2 + \phi^2$ is the lift of $D_{M \times S^1_\beta}^2 + \phi^2$ and $\left(D_{\widetilde{M} \times S^1_\beta}^2 + \phi^2\right) \otimes 1_N$ is the lift of $D_{M \times S^1_\beta,\alpha}^2 + \phi^2$,  Theorems~\ref{pthm2} and~\ref{pthm3}, \emph{mutatis mutandis}, therefore imply the following result:

\begin{prop}
Let $\Gamma \to \tilde{M} \to M$, $\tilde{\cV} \to X$, $\tilde{D}$ and $\alpha : \Gamma \to U(N)$ be as above, and define $\cV$, $D$, $\cV_{\alpha}$ and $D_{\alpha}$ as above. Let $V_{\tilde{M}}$, $V_{M}$ and $V_{M,\alpha}$ be defined as above. Then
\begin{equation}
	V_M(\phi) = \frac{1}{\#\Gamma}V_{\widetilde{M}}(\phi) + O(\Lambda^{-\infty}), \quad \text{as $\Lambda \to +\infty$,}
\end{equation}
and
\begin{equation}
	V_{M,\alpha}(\phi) = \frac{N}{\#\Gamma}V_{\widetilde{M}}(\phi) + O(\Lambda^{-\infty}), \quad \text{as $\Lambda \to +\infty$.}
\end{equation}
\end{prop}

This, therefore, explains the factor $\lambda_Y$ in Equations~\ref{VYS3} and~\ref{powerlawP} and Proposition~\ref{amplNGamma}.

\bigskip
\bigskip
\bigskip

{\bf Acknowledgment.} This work is partially supported by NSF grants
DMS-0901221, DMS-1007207, DMS-1201512, PHY-1205440. The second author thanks
Jos\'e Luis Cisneros-Molina for a useful conversation and for pointing out to
us the results of \cite{twist}, and the first author thanks Mathai Varghese for useful conversations.

\end{document}